\documentclass[fleqn,usenatbib]{mnras}

\usepackage{newtxtext,newtxmath}

\usepackage[T1]{fontenc}
\usepackage{ae,aecompl}
\usepackage{soul}

\usepackage{graphicx}	
\usepackage{amsmath}	
\usepackage[usenames]{xcolor}
\usepackage{natbib}
\usepackage{hyperref}
\usepackage{xfrac}
\usepackage{cases}
\usepackage{changepage}

\def\equationautorefname~#1\null{equation~(#1)}

\DeclareMathAlphabet{\mathpzc}{OT1}{pzc}{m}{it}\definecolor{purple}{RGB}{160,32,240}
\newcommand{\nickk}[1]{\textcolor{black}{ #1}}
\newcommand{\referee}[1]{\textcolor{black}{ #1}}
\newcommand{\nick}[1]{\textcolor{black}{ #1}}
\newcommand{\proofs}[1]{\textcolor{black}{ #1}}

\newcommand{\radmc}{\texttt{radmc3d}}

\newcommand{\Msun}{M_{\odot}}
\newcommand{\Mstar}{M_{\star}}
\newcommand{\Mearth}{m_{\oplus}}
\newcommand{\Mj}{m_{\rm J}}

\newcommand{\Rj}{r_{\rm J}}

\newcommand{\gcm}{\rm g/cm^{2}}
\newcommand{\rb}{r_{\rm Bondi}}

\newcommand{\rshock}{r_{\rm shock}}

\newcommand{\Mp}{m_{\rm p}}

\newcommand{\Tshock}{T_{\rm shock}}

\newcommand{\fdtg}{f_{\rm dust}}

\newcommand{\mdot}{\dot{m}_{\rm p}}

\newcommand{\Fr}{F_{\rm r}}
\newcommand{\acpd}{\alpha_{\rm cpd}}

\title[Protoplanet SEDs]{Spectral Energy Distributions of Disc-Embedded Accreting Protoplanets}

\author[Choksi \& Chiang]{Nick Choksi$^{1 \href{https://orcid.org/0000-0003-0690-1056}{\includegraphics[scale=0.4]{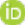}}}$\thanks{E-mail: nchoksi@berkeley.edu} and  Eugene Chiang$^{1,2}$ \\ 
$^{1}$Department of Astronomy, Theoretical Astrophysics Center, and Center for Integrative Planetary Science, University of California, Berkeley, CA 94720, USA\\
$^{2}$Department of Earth and Planetary Science, University of California, Berkeley, CA 94720, USA \\
}

\date{Released \today}
\pubyear{2023}

\begin{document}
\label{firstpage}
\pagerange{\pageref{firstpage}--\pageref{lastpage}}
\maketitle

\begin{abstract} 
Many dozens of
circumstellar discs show signatures of sculpting by planets. To help find these protoplanets by direct imaging, we compute their broadband spectral energy distributions, which overlap with the JWST (James Webb Space Telescope) and ALMA (Atacama Large Millimeter Array) passbands. \referee{We consider how circumplanetary spherical envelopes and circumplanetary discs 
are heated by accretion and irradiation.}
Searches with JWST's NIRCam (Near-Infrared Camera) and the blue portion of MIRI (Mid-Infrared Instrument) are most promising since $\sim$300--1000 K protoplanets outshine their $\sim$20--50 K circumstellar environs at wavelengths of $\sim$2--10 $\mu$m. Detection is easier if  circumplanetary dust settles into discs (more likely for more massive planets) or is less abundant per unit mass gas (because of grain growth or aerodynamic filtration). 
At wavelengths longer than 20 $\mu$m, circumplanetary material is difficult to see against the circumstellar disc's surface layers that directly absorb starlight and reprocess it to the far-infrared. Such contaminating circumstellar emission can be serious even within the evacuated 
gaps observed by ALMA. Only in strongly depleted regions, like the cavity of the transitional disc PDS 70 where two protoplanets have been confirmed, may long-wavelength windows open for protoplanet study. We compile a list of candidate protoplanets and identify those with potentially the highest accretion luminosities, all peaking in the near-infrared.
\end{abstract}



\section{Introduction}
\label{sec:Intro}

For planet formation to mature as an empirical science, it needs to be grounded in observations of forming planets. To date, detections 
of planets accreting from their parent circumstellar discs have been few. The transitional disc system PDS 70 hosts two Jupiter-mass protoplanets, b and c, which radiate primarily in the near-infrared (\citealt{keppler_etal_2018}; \citealt{wang_etal_2020,wang_etal_2021}, and references therein). These objects exhibit millimeter-wave excesses from circumplanetary dust (\citealt{isella_etal_2019, benisty_etal_2021}), and Balmer line and broadband ultraviolet excesses indicating ongoing accretion \citep{haffert_etal_2019, aoyama_ikoma_2019, hashimoto_etal_2020, aoyama_etal_2021, zhou_etal_2021}.
\nickk{Aside from PDS 70b and c, just a handful of potential protoplanets have been imaged, e.g. around AS 209 in the millimetre \citep{bae_etal_2022}, and around MWC 758, AB Aur, and HD 169142 in the near-infrared (\citealt{wagner_etal_2023}; \citealt{currie_etal_2022}; \citealt{zhou_etal_2022, zhou_etal_2023}; \citealt{hammond_etal_2023}). } So far these planet candidates lack the multi-wavelength coverage needed to strengthen their candidacy and determine whether they are still in the process of forming (read: accreting). Helping to fill in the panchromatic picture is a goal of this paper.

\nick{Why aren't there more 
confirmed accreting protoplanets? It is not for want of candidate protoplanetary discs. Dozens of discs with morphological or kinematic signatures suggestive of gravitational forcing by planets have been searched \citep[e.g.][]{zurlo_etal_2020, asensio-torres_etal_2021, andrews_etal_2021, cugno_etal_2023b, cugno_etal_2023}. } Are we looking for protoplanets at the right wavelengths? How important is dust extinction? What about contaminating light from the surrounding circumstellar disc?

We address these questions by
modeling the spectral energy distributions (SEDs) of accreting protoplanets and their environments. Early SED calculations considered 
simple power-law circumplanetary discs and ignored background circumstellar material (e.g.~\citealt{zhu_2015,eisner_2015}). 
\nick{More recent attempts are analytic and consider 
circumplanetary discs surrounded by spherical envelopes with varying density profiles (\citealt{adams_batygin_2022, taylor_adams_2024}), or
are numerical and rely on 3D radiative transfer and/or hydrodynamic simulations 
(\citealt{szulagyi_etal_sed1, szulagyi_etal_sed2, szulagyi_etal_sed4, krieger_wolf_2022}).} We adopt an intermediate, semi-analytic approach that tries to balance realism with efficient exploration of parameter space and easy physical interpretation.

\referee{Flows around planets can be more spherical or disc-like depending on how energy and angular momentum are transported (\citealt{szulagyi_etal_2016}; \citealt{fung_etal_2019}; \citealt{schulik_etal_2020}; \citealt{choksi_etal_2023}; \citealt{krapp_etal_2024}).} We consider both geometries in simplified limits, and calibrate our models against 
hydrodynamic 
simulations (\citealt{choksi_etal_2023}, \citealt{li_etal_2023}). For the most part we concentrate on the parameter space inspired by the DSHARP survey of circumstellar discs \citep{andrews_etal_2018,huang_etal_2018} and their candidate protoplanets (\citealt{zhang_etal_2018}): super-Earths to sub-Jupiters ($\sim$10-100 $m_\oplus$) embedded in low-density circumstellar gaps at orbital distances of $\sim$100 au. 
\referee{We consider how circumplanetary material reprocesses shorter wavelength radiation generated in the planetary accretion shock to longer wavelengths, as well as the radiation generated internally in the accretion flow by viscous dissipation and compression.} We model only the broadband SED redward of the optical (not the ultraviolet excess or line emission from the accretion shock; cf.~\citealt{aoyama_etal_2021} and references therein), with an eye toward assessing detection prospects for the Atacama Large Millimeter Array (ALMA) and the James Webb Space Telescope (JWST).

Section \ref{sec:model} describes how we compute SEDs in spherical and disc geometries. Section \ref{sec:results} presents our results across parameter space, and provides a test application to PDS 70c. We summarize and discuss in section \ref{sec:summary}.

\begin{figure*} 
\includegraphics[width=0.99\columnwidth]{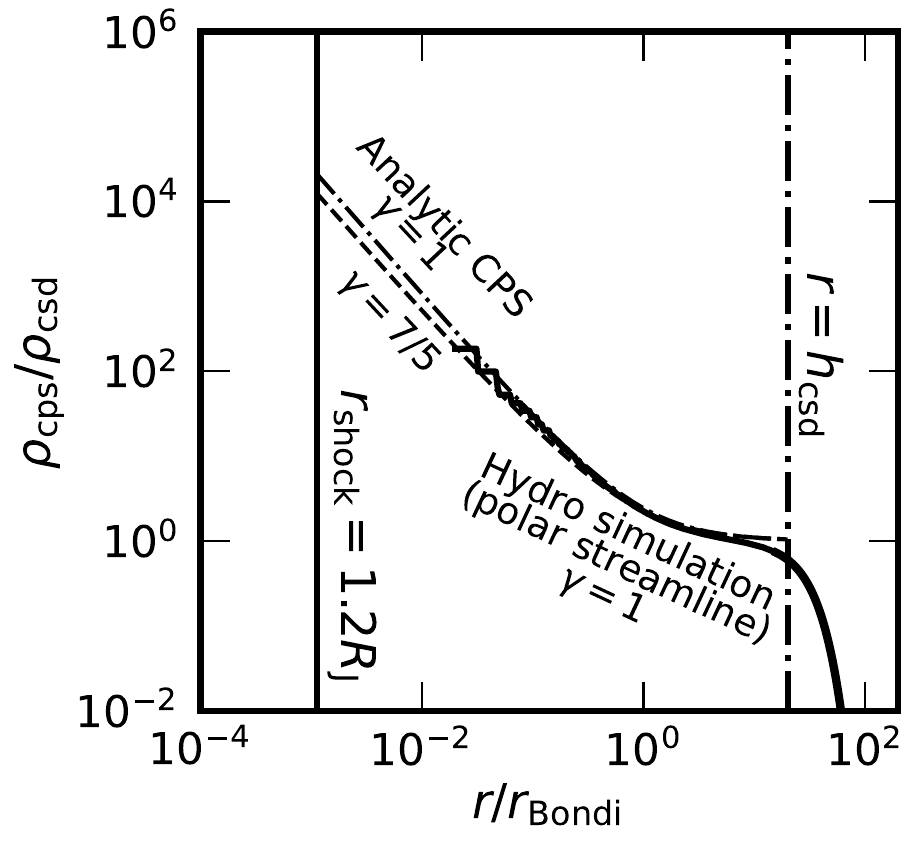}
\caption{ \referee{Gas density $\rho_{\rm cps}$ at distance $r$ from the planet in a spherically symmetric CPS (circumplanetary sphere) model. The dashed and dot-dash curves show Bondi density profiles for two values of the adiabatic index $\gamma$, extending from the innermost CPS radius $\rshock$ (solid vertical line; equation \ref{eqn:rshock} with $F_{\rm r} = 3$) to the outermost CPS radius given by the local circumstellar disc height $h_{\rm csd}$ (vertical dash-dot line).  The solid black curve shows the density from a $\gamma = 1$ 3D-hydro planet-disc simulation for a $17 m_{\oplus}$ planet, measured along a gas streamline directed toward the planet pole \citep{choksi_etal_2023}. The simulation curve extends from $r > h_{\rm csd}$ where $\rho_{\rm cps}$ falls below the circumstellar disc midplane value $\rho_{\rm csd}$, to $\sim$$0.02\,\rb$ at the limit of the simulation's resolution. Where it can, the curve from the hydro simulation supports the Bondi solution as a description of the gas flowing into the planet's gravitational sphere of influence.
Since $\gamma$ hardly affects the density profile, we arbitrarily use the $\gamma = 7/5$ curve when evaluating $\rho_{\rm cps}$ (while not using the temperature adiabat for $\gamma = 7/5$; see section \ref{subsec:cps} for how we compute the CPS temperature $T_{\rm cps}$).  Using the Bondi infall solution $\rho_{\rm cps}(r)$ helps us capture the region near $\rshock$ which dominates the optical depth ($\rho_{\rm cps} r \propto r^{-1/2}$), and which can be difficult to resolve in numerical simulations \citep[cf.][]{szulagyi_etal_sed1, szulagyi_etal_sed2}.
}   }
  \label{fig:denprofile}
\end{figure*}

\begin{figure*} 
\includegraphics[width=0.99\columnwidth]{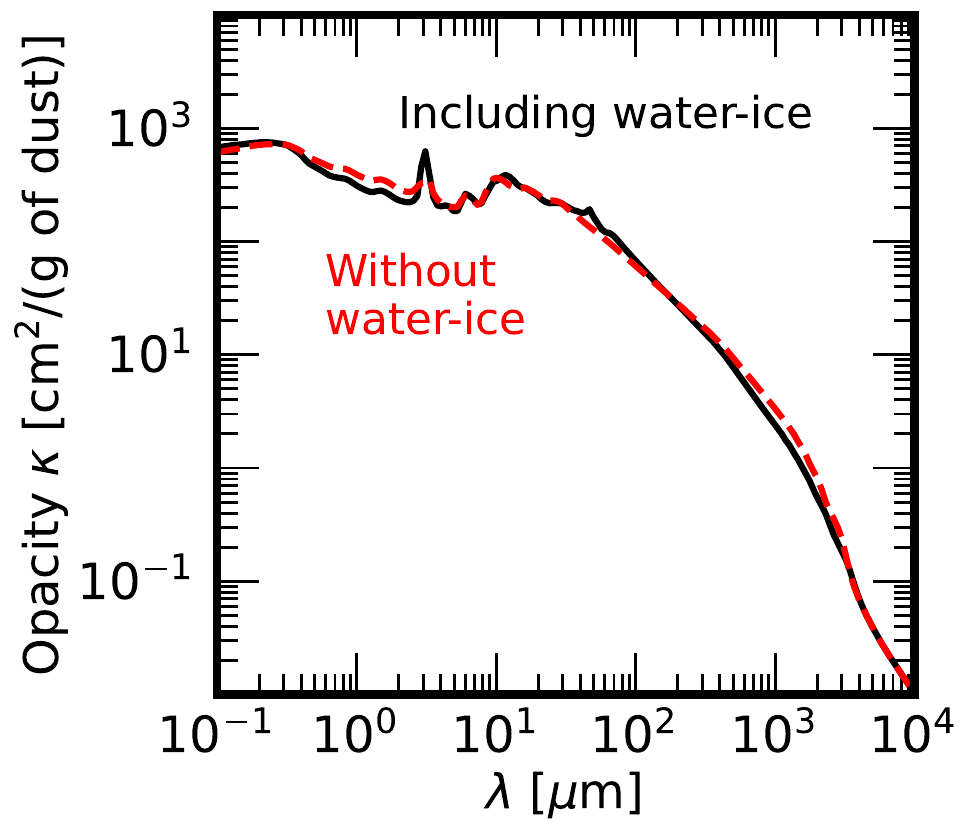}
\caption{ Dust absorption opacities $\kappa$ used in our SED models as a function of wavelength $\lambda$. These are computed using the \texttt{dsharp-opac} package and have been averaged over the fiducial grain size distribution used by the DSHARP collaboration \citep{birnstiel_etal_2018}. The solid black and dashed red curves plot opacities for grains with and without water ice, respectively. A vibrational mode of water ice boosts opacities around $\lambda = $ 2.7 $\mu$m. By default, we compute SEDs including water ice. We check this assumption post-facto by measuring from $\radmc$ the dust temperature where photons of wavelength $\lambda = 2.7$ $\mu$m reach radial optical depth unity (starting from outside the CPS and integrating toward the centre). If this temperature exceeds the ice sublimation temperature of 150 K, we rerun $\radmc$ using the opacity curve excluding water ice.
}
\label{fig:opacity}
\end{figure*}

\begin{table*}
\begin{tabular}{ccc}
\hline 
\\[-2mm]
Parameter Name &  Description & Value(s) \\ 
\hline 
$\Mstar$ & Host star mass & $\Msun$ \\ 
$T_{\star}$ & Host star effective temperature & 4500 K \\ 
$R_{\star}$ & Host star radius & 2.5$R_{\odot}$ \\ 
$t_{\rm age}$ & System age & 1\,$\mathrm{Myr}$ \\ 
$a$ & Protoplanet orbital radius & \{100,\,300\} au \\ 
$h_{\rm csd}/a$ & Circumstellar disc aspect ratio at 100 au & 0.1 \\ 
$\Sigma_{\rm csd}$ & Circumstellar disc gas surface density at $a$ & \{$0.03$,\,$0.3$,\,$3$\}\, $\gcm$ \\ 
$\fdtg$ & Dust-to-gas ratio & $\{10^{-3},\,10^{-2}\}$
\\
$m_{\rm p}$  & Protoplanet mass & $\{30, \,100, \,300\}\Mearth$ \\ 
$F_{\rm r}$   & Protoplanet radius, normalised by \cite{thorngren_etal_2019} mass-radius relation 
& $\{3, \,10\}$ \\ 
$\alpha_{\rm cpd}$ & Shakura-Sunyaev viscosity parameter in CPD  &  $\{10^{-4},\,10^{-3},\,10^{-2}\}$ \\ 
\\[-2mm]
\hline 
\hline 
\end{tabular}
\caption{ Summary of parameters in our SED models. 
}
\label{tab:summary}
\end{table*}

\section{SED Model}
\label{sec:model}
We consider a planet orbiting a star of mass $\Mstar = \Msun$, effective temperature $T_{\star} = 4500$ K, and radius $R_{\star} = 2.5R_{\odot}$. These stellar properties derive from evolutionary tracks evaluated at a system age $t_{\rm age} = 1\,\mathrm{Myr}$ \citep{choi_etal_2016}. The planet is on a circular orbit with radius $a = 100$ au in the midplane of a circumstellar disc (CSD).
The local 
CSD height is $h_{\rm csd} = 0.1a$, the
sound speed $c_{\rm csd} = h_{\rm csd}\Omega$, and the midplane
temperature $T_{\rm csd} = \mu m_{\rm H} c_{\rm csd}^2/k_{\rm B} = 26$ K,
where $\Omega = \sqrt{G\Mstar/a^3}$ is the Keplerian orbital frequency,
$G$ is the gravitational constant, $k_{\rm B}$ is Boltzmann's
constant, $\mu = 2.4$ is the gas mean molecular weight, and $m_{\rm H}$ is the mass of the hydrogen atom.  The local
surface density (inside whatever gap the planet may have opened) is
assumed to be $\Sigma_{\rm csd} = 0.3\,\rm \gcm$, and the volume
density at the midplane is
$\rho_{\rm csd} = \Sigma_{\rm csd}/\left(\sqrt{2\pi}h_{\rm
    csd}\right)$.
These fiducial parameters are broadly consistent with observations of DSHARP gaps (e.g.~\citealt{choksi_chiang_2022, choksi_etal_2023}, and references therein). Alternate choices are listed in Table \ref{tab:summary} and tested in Section \ref{subsec:params}.

We calculate SEDs of accreting protoplanets of mass
$m_{\rm p} = \{30,\,100,\,300\}\,\Mearth$
(cf. \citealt{szulagyi_etal_sed1} who focus on more massive
planets). Sub-Jupiter masses are typically indicated by the widths of
observed disc gaps \citep{dong_fung_2017, zhang_etal_2018}. Our
parameters are such that the disc height $h_{\rm csd}$ is 
greater than or equal to the planet's Bondi radius
\begin{align} \label{eqn:bondi}
\rb = \frac{G\Mp}{c_{\rm csd}^2}.
\end{align}
In this ``subthermal'' regime, the planet's gravitational sphere of influence is its Bondi sphere (not its Hill sphere) and we can neglect the vertical density gradient of the background disc (e.g. \citealt{choksi_etal_2023}). 
Most suspected gap-opening planets from DSHARP are subthermal (table 1 of \citealt{choksi_etal_2023}).

The planet's mass accretion rate is assumed to be its time-averaged rate over its age:
\begin{equation} \label{eqn:mdot}
\dot{m}_{\rm p} = \Mp/t_{\rm age}.
\end{equation}
\nickk{Planets growing much faster than (2) are unlikely to be observed because episodes of fast accretion would be short-lived \citep[e.g.][]{choksi_chiang_2022}.}

\nick{We assume the planet has entered the hydrodynamic ``runaway'' phase of growth, during which flows can accelerate to supersonic speeds (e.g. \citealt{ginzburg_chiang_2019a}, and references therein).}
At a distance $\rshock$ from the planet's centre, the accreting gas passes through a shock into the planet's (nearly hydrostatic) atmosphere, converting its bulk kinetic energy into heat. The post-shock emission is assumed to be from a blackbody of luminosity and temperature
\begin{align}
L_{\rm acc} &= \frac{G m_{\rm p} \mdot}{\rshock} \label{eqn:L}   \\ 
\Tshock &= \left(\frac{L_{\rm acc}}{4 \pi \rshock^2 \sigma_{\rm SB}}\right)^{1/4} 
\label{eqn:blackbody}
\end{align}
where $\sigma_{\rm SB}$ is the Stefan-Boltzmann constant. \nickk{Equation (\ref{eqn:blackbody}) agrees quantitatively with radiation-hydrodynamic simulations of accretion shocks that find $\gtrsim$ 80\% of the accretion power thermalises immediately, for the sub-Jupiter masses that we focus on   (\citealt{aoyama_etal_2020}, their fig. 10 and equations A3-A5; see also \citealt{marleau_etal_2017}). Note that standard ``cold start'' or ``hot start'' cooling models of planets passively radiating into empty space do not apply for the actively accreting protoplanets of interest here (\citealt{berardo_etal_2017}; cf. \citealt{zhu_2015, szulagyi_etal_sed1}). We do not explicitly model the $\lesssim$ 20\% of the accretion power that is released in non-thermal 
emission (mostly in Lyman-$\alpha$; \citealt{aoyama_etal_2020, aoyama_etal_2021}). Such radiation would be thermalised if the planet's surroundings (at $r > \rshock$) are optically thick, as they often are for our models.}

The protoplanet's size is essentially given by $\rshock$. We set
\begin{align}
\frac{\rshock}{\Rj} = \Fr  \left[ 0.96 + 0.21 \log_{10}\frac{\Mp}{\Mj} - 0.2\left(\log_{10}\frac{\Mp}{\Mj}\right)^2\right]
\label{eqn:rshock}
\end{align}
where $r_{\rm J}$ and $m_{\rm J}$ are the radius and mass of Jupiter. 
The expression in square brackets is a mass-radius relation fit to gas giants that are Gyrs old \citep{thorngren_etal_2019}. We introduce the parameter $F_{\rm r} > 1$ because protoplanets may not have finished contracting. \nickk{We test $F_{\rm r} = \{3,\,10\}$. These values are similar to those inferred for the PDS 70 protoplanets based on their near-infrared SEDs (\citealt{muller_etal_2018, wang_etal_2020}; our section 3.4), and lie within the range $F_{\rm r} \approx 2-40$ predicted by protoplanet cooling models for planets aged 1-10 Myr (\citealt{ginzburg_chiang_2019b}, their figs. 2 and 4; the quoted range accounts for uncertainty in the disc viscosity and atmosphere opacity).}


\referee{A fraction of the blackbody radiation from the accretion shock
is re-processed by dust exterior to $r_{\rm shock}$. The emergent SED includes whatever planetary radiation is not intercepted, the re-processed emission from dust, and radiation generated by compressional and dissipative heating in the accretion flow. We try to be agnostic about the geometry of circumplanetary material by computing SEDs for both circumplanetary spheres (CPSs) and circumplanetary discs (CPDs, section \ref{subsec:cpd}; the CPD models actually combine discs + spheres). For spherical geometries, we use the Monte Carlo radiative transfer code $\radmc$
\citep{dullemond_etal_2012}, accounting for additional heating from compression during infall. For disc geometries, we adapt the two-layer irradiated disc model of \citet{chiang_goldreich_1997}, accounting for additional heating from viscous dissipation.  The challenge for observers is to measure the protoplanet's SED against the spectrum of re-processed starlight emitted by the local circumstellar disc (CSD); the latter is computed in section \ref{subsec:csd}.}

\subsection{Circumplanetary sphere (CPS) models}
\label{subsec:cps}

\referee{Over radial distances $r=\rshock$ to $r=h_{\rm csd}$ measured from the planet centre, the gas density $\rho_{\rm cps}$ is taken from a spherically symmetric Bondi solution (see Appendix \ref{appendix:bondi} for equations solved). 
Our density profile $\rho_{\rm cps}(r)$ corresponds to mass inflow at the Bondi rate
\begin{equation}
  \dot{m}_{\rm in} = \pi q(\gamma) G^2\Mp^2 \rho_{\rm csd}/c_{\rm csd}^3 \label{eqn:mdot_in}
\end{equation}
where $\gamma$ is the adibatic index and $q(\gamma) = \left[ 2 / (5-3\gamma)\right]^{ (5 - 3\gamma)/(2\gamma - 2)}$. Hydrodynamic simulations of subthermal planets embedded in 3D disks support this Bondi infall density profile, at least for $\gamma = 1$; see Figure \ref{fig:denprofile}, which shows how the isothermal density profile reproduces data from a 3D, global, isothermal planet-disk simulation by Choksi et al. (\citeyear{choksi_etal_2023}; see also \citealt{dangelo_etal_2003}; \citealt{machida_etal_2010}; \citealt{li_etal_2023}; for adiabatic 3D simulations, see \citealt{fung_etal_2019}). 
Figure \ref{fig:denprofile} also plots $\rho_{\rm cps}(r)$ for $\gamma = 7/5$, showing that the value of $\gamma$ hardly affects the CPS density profile, which follows $\rho_{\rm cps} \propto r^{-3/2}$ when $r \ll \rb$ (free-fall) and $\rho_{\rm cps} \simeq \rho_{\rm csd}$ in the opposite limit, regardless of $\gamma$.}

\referee{In this paper, we use the density profile for $\gamma = 7/5$ and $\dot{m}_{\rm in}$ given by equation (\ref{eqn:mdot_in}), but note that (i) the planet's {\it net} mass accretion rate $\dot{m}_{\rm p}$ does not equal $\dot{m}_{\rm in}$, and (ii) the  temperature profile does not correspond to a $\gamma = 7/5$ adiabat.
Regarding (i): our model distinguishes between the inflow rate $\dot{m}_{\rm in}$ (given by
equation \ref{eqn:mdot_in}) and the net planetary accretion rate
$\dot{m}_{\rm p} \equiv \dot{m}_{\rm in} - \dot{m}_{\rm out}$ (given
by equation \ref{eqn:mdot}). In principle, much of the gas entering a
planet's gravitational sphere of influence ($\dot{m}_{\rm in}$,
dominated by polar streamlines) can also exit it ($\dot{m}_{\rm out}$,
dominated by equatorial streamlines; \citealt{ormel_etal_2015,
  cimerman_etal_2017, bethune_rafikov_2019, choksi_etal_2023}).
  For our fiducial parameters, $\mdot = \Mp/t_{\rm age}$ is smaller
than $\dot{m}_{\rm in}$ by factors of 10-100 (see also fig. 13 of
\citealt{choksi_etal_2023}). We will also explore the limiting case $\dot{m}_{\rm p} = \dot{m}_{\rm in}$, which appears to apply to PDS 70bc.}


\referee{Regarding (ii): our CPS temperature profile accounts for both radiative heating from (passive reprocessing of) thermal photons emitted by the planet's accretion shock, and compressional heating of gas during infall. The temperature fields considering each heat source in isolation are 
$T_{\rm rad}(r)$ and $T_{\rm compress}(r)$, respectively. The net CPS temperature is given by
\begin{equation}
T_{\rm cps}^4=  T_{\rm rad}^4 + T_{\rm compress}^4 + T_{\rm csd}^4 
\end{equation}
where the last term is a temperature floor set by heating from the star and background circumstellar disc. The net temperature $T_{\rm cps}$ is fed into the \texttt{sed} routine of $\radmc$ to obtain the emergent SED $\lambda L_{\lambda} \equiv 4\pi d^2\lambda F_{\lambda}$, where $d$ is distance to the source, $\lambda$ is photon wavelength, and $F_{\lambda}$ is flux per unit wavelength. When computing the CPS SED, we exclude cells whose temperature is within 20\% of the background $T_{\rm csd}$. The CSD's emission is separately accounted for in section \ref{subsec:csd}.}

Dust and gas are assumed uniformly mixed throughout the CPS (cf.~\citealt{krapp_etal_2022, krapp_etal_2024} who consider non-uniform mixtures). 
We explore a solar dust-to-gas ratio of $\fdtg = 10^{-2}$, as well as a subsolar ratio of $10^{-3}$ since dust may be filtered out of planet-carved gaps \citep[e.g.][]{dong_etal_2017}. We assume dust grains follow the fiducial size distribution adopted by the DSHARP collaboration (orange curve in fig. 5 of \citealt{birnstiel_etal_2018}), and calculate dust absorption opacities using the \texttt{dsharp-opac} package.\footnote{\url{https://github.com/birnstiel/dsharp_opac}} Figure \ref{fig:opacity} plots the dust opacity $\kappa$ vs.~wavelength $\lambda$ for grains with and without water ice. The main effect of including water ice is to boost the opacity around $\lambda = $ 2.7 $\mu$m. By default, we compute SEDs using the opacity curve that includes water ice, but omit water ice if temperatures prove too high (see caption to Fig.~\ref{fig:opacity} for details).



\referee{To calculate the temperature $T_{\rm rad}(r)$ that the CPS would attain if  heated only by planetary radiation, we use $\radmc$'s $\texttt{mctherm}$ routine. The planet is modeled as a point source with the blackbody SED defined by equations (\ref{eqn:L}) and (\ref{eqn:blackbody}). The planet source emits $10^4$ photon ``packages'' across 30 wavelength bins uniformly distributed in log-space between 0.1 $\mu$m and 10$^4$$\mu$m. Radiative transfer is performed in spherical coordinates $(r,\,\theta,\,\psi)$ centred on the planet, for radius $r$ and polar and azimuthal angles $\theta$ and $\psi$, respectively. The number of grid cells is $\{N_{\rm r},\,N_{\theta},\,N_{\psi}\} = \{64,1,1\}$ (assuming spherical symmetry). The spacing between neighboring radial cells grows in proportion to the distance from the planet according to $r_{i+1}/r_i - 1 = 1/3$, where $r_i$ is the inner edge of the $i$th cell. }

\referee{The temperature $T_{\rm compress}(r)$ to which gas is heated by compression during infall is determined by setting the local radiative cooling time
\begin{align}
t_{\rm cool} &= \frac{r \rho_{\rm cps} k_{\rm B}T }{\mu m_{\rm H}\sigma_{\rm SB}T^4\langle \tau \rangle_{T}}\,\,&\mathrm{if} \,\,\langle \tau \rangle_{T} < 1\ \nonumber \\ 
&= \frac{r \rho_{\rm cps} k_{\rm B}T }{\mu m_{\rm H} \sigma_{\rm SB}T^4/\langle \tau \rangle_{T}}\,\,&\mathrm{if} \,\,\langle \tau \rangle_{T} \geq 1 
\label{eqn:tcool}
\end{align}
equal to the local infall time (i.e.~compressional heating time) $t_{\rm dyn} = r/\sqrt{Gm_{\rm p}/r}$. Equation (\ref{eqn:tcool}) is an order-of-magnitude estimate for the cooling time that accounts for optically thin or thick conditions, where $\langle \tau \rangle_{T} = \langle \kappa \rangle_{T}  \rho_{\rm cps} f_{\rm dust}r$ is the Planck-averaged optical depth at the local $T$. The rationale behind setting $t_{\rm cool} = t_{\rm dyn}$ is that if $t_{\rm cool} \ll t_{\rm dyn}$ the gas would be isothermal, whereas if $t_{\rm cool} \gg t_{\rm dyn}$ the gas would be adiabatic. We have shown that neither limiting case is self-consistent (see Fig.~\ref{fig:cps_summary} below), and so the actual temperature profile must be intermediate, set by $t_{\rm cool} = t_{\rm dyn}$. We solve for the temperature field $T_{\rm compress}(r)$ that satisfies this condition (by numerical iteration since $\langle \kappa \rangle_{T}$ is computed numerically). Since $t_{\rm cool} \propto 1/T^3$, $T_{\rm compress}$ is a stable equilibrium.}

Our main results for CPS models are presented without including gas opacity or dust scattering. \nick{In appendix \ref{appendix:scat} we show how including dust scattering in the CPS does not much alter the CPS SED.} For an assessment of how dust scattering may affect the light from the circumstellar disc (CSD) and by extension protoplanet detectability, see the end of section \ref{subsec:csd}.

\subsection{Circumplanetary disc (CPD) models}
\label{subsec:cpd}
We model CPDs that are heated both internally by viscous dissipation,
and externally by radiation from the planet and the star
(cf. \citealt{zhu_2015} who consider only viscous heating). The CPD is assumed to be truncated by the planet's magnetosphere at an inner radius $r_{\rm in} = 2r_{\rm shock}$, characteristic of accreting gas giants whose magnetic fields are in energy equipartition with their convective flows (section 2 and bottom panel of fig.~1 of \citealt{ginzburg_chiang_braking}; see also \citealt{batygin_2018}, who obtain similar values). The CPD extends to an outer radius $r_{\rm out}$ where by definition the CPD interior temperature reaches the background circumstellar disc temperature $T_{\rm csd}$. At $r > r_{\rm out}$, the dust distribution reverts to a spherically CPS.



The CPD has the steady-state surface density
\begin{equation}
\Sigma_{\rm cpd} = \frac{\mdot}{3\pi \acpd c_{\rm cpd}h_{\rm cpd}} 
\label{eqn:mdot_visc}
\end{equation}
where $\acpd = \{10^{-4},\,10^{-3},\,10^{-2}\}$ is the \cite{shakura_sunyaev_1973} viscosity parameter, $c_{\rm cpd} = \sqrt{k_{\rm B}T_{\rm cpd,mid}/(\mu m_{\rm H})}$ is the sound speed at the midplane temperature $T_{\rm cpd,mid}$, and $h_{\rm cpd} = c_{\rm cpd}/\sqrt{G\Mp/r^3}$ is the disc's vertical height in hydrostatic equilibrium with the planet's gravity. The corresponding midplane density is $\rho_{\rm cpd,mid} = \Sigma_{\rm cpd}/(\sqrt{2\pi}h_{\rm cpd})$.

We assume the energy released by viscous dissipation is transported out by radiative diffusion, so that the midplane is heated to 
\begin{equation}
T_{\rm cpd,mid} = \mathrm{min}\left[\left(\frac{3 G\Mp \mdot \langle \tau_{\rm cpd}\rangle_{T_{\rm cpd,mid}} }{16\pi \sigma_{\rm SB} r^3  }  \right)^{1/4},\,1500\,\mathrm{K}\right]
\label{eqn:Tmid}
\end{equation}
where $\langle \tau_{\rm cpd} \rangle_{T_{\rm cpd,mid}} = \langle \kappa\rangle_{T_{\rm cpd,\,mid}}\Sigma_{\rm cpd}f_{\rm dust}$ and $\langle \kappa \rangle_T$ is the dust-only opacity averaged over the Planck function at temperature $T$. 
The temperature ceiling at 1500 K accounts for how dust sublimation thermostats the midplane temperature \citep{dalessio_etal_1998, dalessio_etal_1999}. Using the wavelength-dependent \texttt{dsharp-opac} opacities excluding water ice (section \ref{subsec:cps}), we numerically solve equations (\ref{eqn:mdot_visc}) and (\ref{eqn:Tmid}) for $\Sigma_{\rm cpd}(r)$, $T_{\rm cpd,\,mid}(r)$, and $h_{\rm cpd}(r)$, and verify post-facto our assumption that the CPD is optically thick. For reference, when the opacity is constant and dust has not sublimated, $\Sigma_{\rm cpd} \propto r^{-3/5}$, $T_{\rm cpd} \propto r^{-9/10}$, and $h_{\rm cpd} \propto r^{21/20}$. The last scaling implies the CPD is flared but only barely so. 

The portion of the CPD energy budget from passive reprocessing is computed following \cite{chiang_goldreich_1997} and \cite{chiang_etal_2001}. Grains in the disk surface layers are directly exposed to radiation from the planet and attain the equilibrium temperature
\begin{equation}
T_{\rm cpd,s} = \Tshock \left(\frac{\rshock}{r}\right)^{1/2}\left(\frac{\langle \kappa \rangle_{T_{\rm shock} } }{8 \langle \kappa \rangle_{ T_{\rm cpd,s} }} \right)^{1/4}.
\label{eqn:Ts}
\end{equation}
Half the radiation emitted by the surface grains is directed away from the CPD to be reprocessed by the surrounding CPS. The other half is directed into the CPD interior, heating it to a minimum temperature
\begin{equation}
T'_{\rm cpd,i} = \Tshock \left(\frac{\rshock}{r}\right)^{1/2}\left(\frac{\gamma_{\rm cpd}}{4}\right)^{1/4} 
\label{eqn:Ti}
\end{equation}
which neglects the contribution from viscous heating (to be added below), and which assumes the interior is optically thick. Here
\begin{equation}
\gamma_{\rm cpd} = rd(H_{\rm cpd}/r)/dr + 0.4\rshock/r
\end{equation}
is the angle at which radiation from the planet grazes the CPD photosphere at height $H_{\rm cpd}$ above the midplane. The ratio $H_{\rm cpd}/h_{\rm cpd}$ is a logarithm on the order of a few, and we fix it to be 3. Note the irradiation grazing angle $\gamma_{\rm cpd}$ is assumed to depend on $T_{\rm cpd,mid}$ which is set by viscous heating, rather than $T'_{\rm cpd,i}$; this assumption is safe since $T_{\rm cpd,mid}$ is the highest disc temperature in our models. 

Equation (\ref{eqn:Ti}) accounts only for passive reprocessing and neglects viscous heating. Accounting only for viscous heating gives an effective (photospheric) temperature for the CPD of $T_{\rm cpd,mid}/\left(\langle \tau_{\rm cpd} \rangle_{T_{\rm cpd,mid}} /2\right)^{1/4}$. We combine these heating terms to estimate a net interior temperature $T_{\rm cpd,i}$:
\begin{align}
T_{\rm cpd,i}^4 = (T'_{\rm cpd,i})^4 + \frac{T_{\rm cpd,mid}^4}{\langle \tau_{\rm cpd} \rangle_{T_{\rm cpd,mid}} /2}
\end{align}
which ensures energy flux conservation.

The SED for a CPD viewed face-on is computed as 
\begin{align} \label{eqn:cpd_sed}
\left( \lambda L_\lambda \right)_{\rm cpd} = 8\pi^2 \lambda \int_{r_{\rm in} }^{r_{\rm out}} I_\lambda(r) r dr 
\end{align}
where $I_\lambda$ is the intensity per unit wavelength emitted from disc radius $r$. The contributions to $I_\lambda$ are from the optically thick disc interior and the optically thin surface:
\begin{align}
I_\lambda =  B_{\lambda}(T_{\rm cpd,i}) +  B_{\lambda}(T_{\rm cpd,s})  
\frac{\gamma_{\rm cpd} \kappa(\lambda)}
{\langle \kappa \rangle_{ T_{\rm shock} }} 
\label{eqn:cpd_contribute}
\end{align}
where $B_{\lambda}(T)$ is the Planck function evaluated at temperature $T$, and the factor multiplying $B_\lambda(T_{\rm cpd,s})$ is the optical depth of the optically thin surface layer. For $\kappa(\lambda)$, we use opacities that include water ice if $T_{\rm cpd, s} < 150$ K, and otherwise do not. We define $r = r_{\rm out}$ to be the location where $T_{\rm cpd,i}$ (the lowest computed temperature of the disc) crosses the system floor temperature of $T_{\rm csd}$.


To the CPD SED we add the planet SED:
\begin{align} \label{eqn:cpd+planet}
\left( \lambda L_\lambda \right)_{\rm planet+cpd} = 4\pi^2 \rshock^2 \lambda B_{\lambda}(T_{\rm shock}) +\left( \lambda L_\lambda \right)_{\rm cpd} .
\end{align}
The emission from the planet+CPD is re-processed by a CPS dust shell from $r_{\rm out} < r < h_{\rm csd}$. The SED emerging from the shell is computed using $\radmc$ by excising the inner $r < r_{\rm out}$ from a CPS model, and placing at $r=0$ a point source whose SED is given by equation (\ref{eqn:cpd+planet}) for the planet + face-on CPD. The procedure is otherwise the same as in section \ref{subsec:cps}. 
Since in reality the CPD is less bright when viewed edge-on, using the face-on CPD SED to isotropically heat the surrounding CPS overestimates the flux in the emergent SED. The error, however, is only on the order of unity for a shell that is optically thick to CPD radiation, and less for a shell that is optically thin.



\begin{figure*} 
\includegraphics[width=0.99\textwidth]{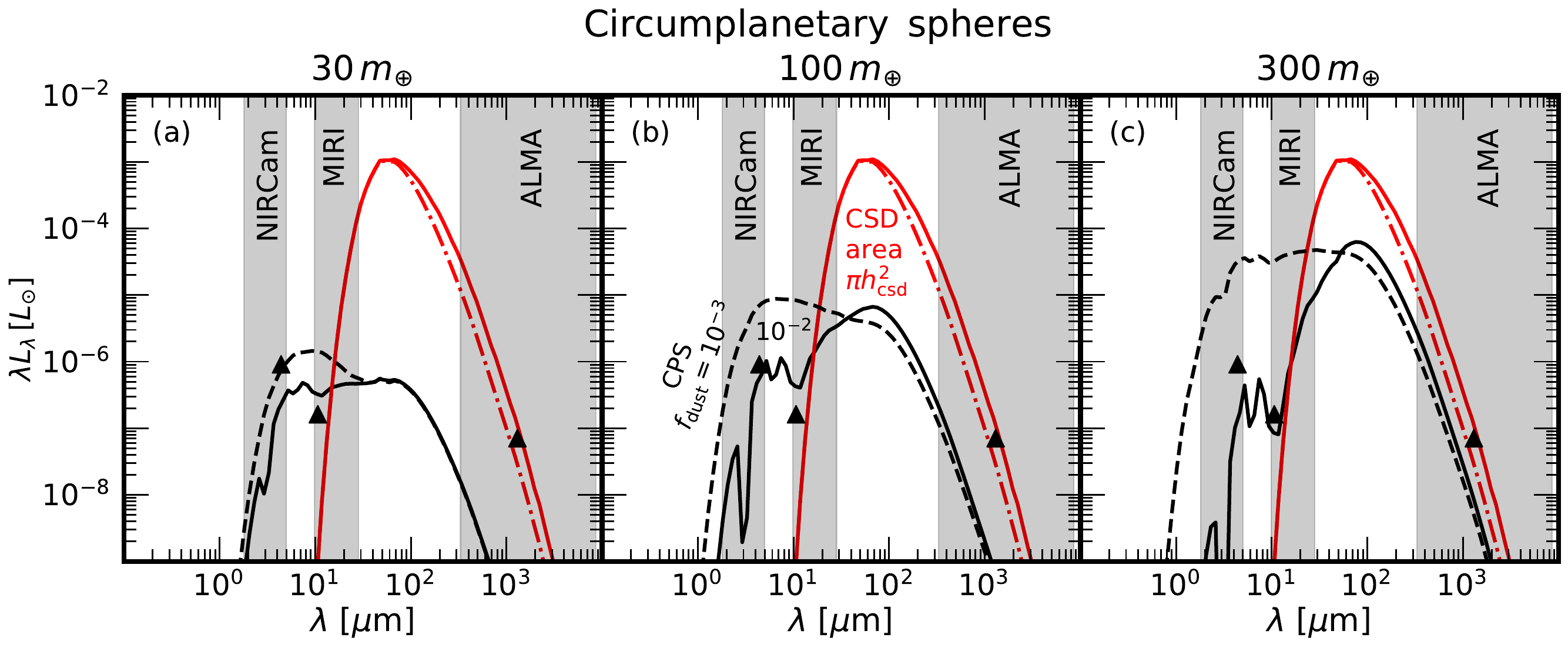}
\caption{Spectral energy distributions (SEDs) of protoplanets surrounded by circumplanetary spheres (CPSs). Different panels correspond to different mass planets as labeled. 
Solid and dashed black curves show the emergent SEDs from CPSs with dust-to-gas ratios of $10^{-2}$ (solar metallicity) and $10^{-3}$, respectively, calculated using $\radmc$. \referee{These SEDs include radiation from the planetary accretion shock that escapes the CPS, plus emission from CPS dust grains heated by the accretion shock and  compressed infalling gas.} Spectral features due to water ice (2.7 $\mu$m) and silicates (10 $\mu$m) are visible in absorption. The solid red curve shows the SED of re-processed starlight from the local circumstellar disc (CSD), scaled to an emitting area $\pi h_{\rm csd}^2$ and plotted for a solar $f_{\rm dust} = 10^{-2}$. Most of the CSD emission is generated by the surface layer (red dash-dot curve) and is insensitive to $f_{\rm dust}$. Shaded regions mark the passbands for coronographic imaging with the James Webb Space Telescope near-infrared camera (JWST NIRCam) and mid-infrared instrument (MIRI), and the wavelength range of the Atacama Large Millimeter Array (ALMA). Protoplanets seem most easily detectable against the local CSD with NIRCam. 
\nick{The upward pointing triangles in the NIRCam and MIRI passbands mark detection limits for coronographic imaging without a background CSD in the F444W and F1065C bands, respectively, at a planet-star separation of 1 arcsecond (5$\sigma$ limit for F444W reported in fig. 5 of \citealt{carter_etal_2023}, and 3$\sigma$ limit for F1065C in fig. 5 of \citealt{boccaletti_etal_2022}; the contrast with the host star is 12.5 and 11.7 magnitudes, respectively).}
The triangle in the ALMA passband corresponds to a 1.3 mm continuum flux of 110 $\mu$Jy at a distance of 100 pc. This is the minimum flux needed 
for a point source to be detected against the background CSD according to injection tests performed on DSHARP gaps
(table 4 of \citealt{andrews_etal_2021}  
for a 90\% recovery fraction).}
  \label{fig:cps_sed}
\end{figure*}

\subsection{Local circumstellar disc (CSD)}
\label{subsec:csd}
We compute the SED of a circular patch of the CSD, of area $\pi h_{\rm csd}^2$,    centred on the planet.  The CSD SED is determined by passive re-processing of starlight and not by viscous dissipation which is typically not competitive with stellar irradiation beyond $a \gtrsim 1$ au \citep{chiang_goldreich_1997}.

We divide the CSD into an interior at temperature $T_{\rm csd}$ and an exposed surface layer at temperature $T_{\rm csd,\,s}$. We set $T_{\rm csd,s} = 50$ K, about twice $T_{\rm csd}$ (e.g.~fig.~5 of \citealt{chiang_etal_2001}). The SED of a patch of area $\pi h_{\rm csd}^2$ is 
\begin{align}
\left(\lambda L_{\lambda}\right)_{\rm csd} &= 4 \pi^2 h_{\rm csd}^2 \lambda \left[ B_{\lambda}(T_{\rm csd})(1 - e^{-\tau_{\rm csd} }) \right. \nonumber  \\
&+ \left. (1 + e^{-\tau_{\rm csd}})B_{\lambda}(T_{\rm csd,\, s}) \frac{ \gamma_{\rm csd} \kappa(\lambda) }{\langle \kappa \rangle_{T_{\star}}}   \right],
\label{eqn:csd_sed}
\end{align}
where $\tau_{\rm csd}(\lambda) \hspace{-1mm} = \kappa(\lambda)\Sigma_{\rm csd}f_{\rm dust}$, and
 $\gamma_{\rm csd} = 0.4R_{\star}/a + ad(H_{\rm csd}/a)/da$ is the angle at which stellar radiation strikes the CSD photosphere at height $H_{\rm csd}$ above the midplane. We take $H_{\rm csd} = 3h_{\rm csd}$ and $h_{\rm csd} \propto a^{2/7}$ \citep{chiang_goldreich_1997, chiang_etal_2001}; together these yield $\gamma_{\rm csd} \simeq (6/7)h_{\rm csd}/a$. Because $T_{\rm csd,s} < 150$ K, we use opacities that include water ice. Note that we have not assumed the CSD interior to be optically thick.

Scattered starlight from the CSD is not explicitly modeled but we assess its significance here. For a blackbody star of luminosity $L_\star$ whose SED peaks at wavelength $\lambda_\star$, we estimate the starlight scattered off an area $\pi h_{\rm csd}^2$ of the CSD surface as
\begin{align}
\left(\lambda L_{\lambda}\right)_{\rm csd,scat} \sim \frac{L_\star (\lambda_\star/\lambda)^3}{4\pi a^2} \, \gamma_{\rm csd} \, \pi h_{\rm csd}^2 \,Q_{\rm scat}(\lambda)
\end{align}
where $L_\star (\lambda_\star/\lambda)^3$ is the broadband Rayleigh-Jeans stellar luminosity at wavelength $\lambda$, and $Q_{\rm scat}$ is the dust scattering efficiency or albedo. For $L_\star \sim L_\odot$, $\lambda_\star \sim 1 \,\mu$m, $\lambda \sim 4 \,\mu$m (corresponding to the F444W passband of JWST's Near-Infrared Camera), and $\gamma_{\rm csd} \sim h_{\rm csd}/a \sim 0.1$, we have $\left(\lambda L_{\lambda}\right)_{\rm csd,scat} \sim 4\times 10^{-6} \, Q_{\rm scat} L_\odot$. 
The scattering albedo $Q_{\rm scat}$ is a sensitive function of dust grain radius $s$ and wavelength; in the Rayleigh limit $2\pi s < \lambda$, $Q_{\rm scat} \sim (2\pi s/\lambda)^4$. For CSD surface grains of size $s = 0.1 \,\mu$m, $\left(\lambda L_{\lambda}\right)_{\rm csd,scat} \sim 2 \times 10^{-9} L_\odot$, well below the $10^{-7}-10^{-4} L_\odot$ emitted thermally by circumplanetary material (depending on planet mass and other parameters; see section \ref{sec:results}). By contrast, for $s = 1 \,\mu$m, scattering may be in the geometric limit and $Q_{\rm scat}$ on the order of unity; then $\left(\lambda L_{\lambda}\right)_{\rm csd,scat} \sim 10^{-6} (Q_{\rm scat}/0.3) L_\odot$ and
scattered near-infrared starlight may overpower circumplanetary emission in some cases. At mid-infrared and longer wavelengths, scattered starlight is unlikely to introduce significant confusion because of the $\lambda^4$ dependence in $Q_{\rm scat}$.

\begin{figure*} 
\vspace{-1.4cm}
\includegraphics[width=0.9\textwidth]{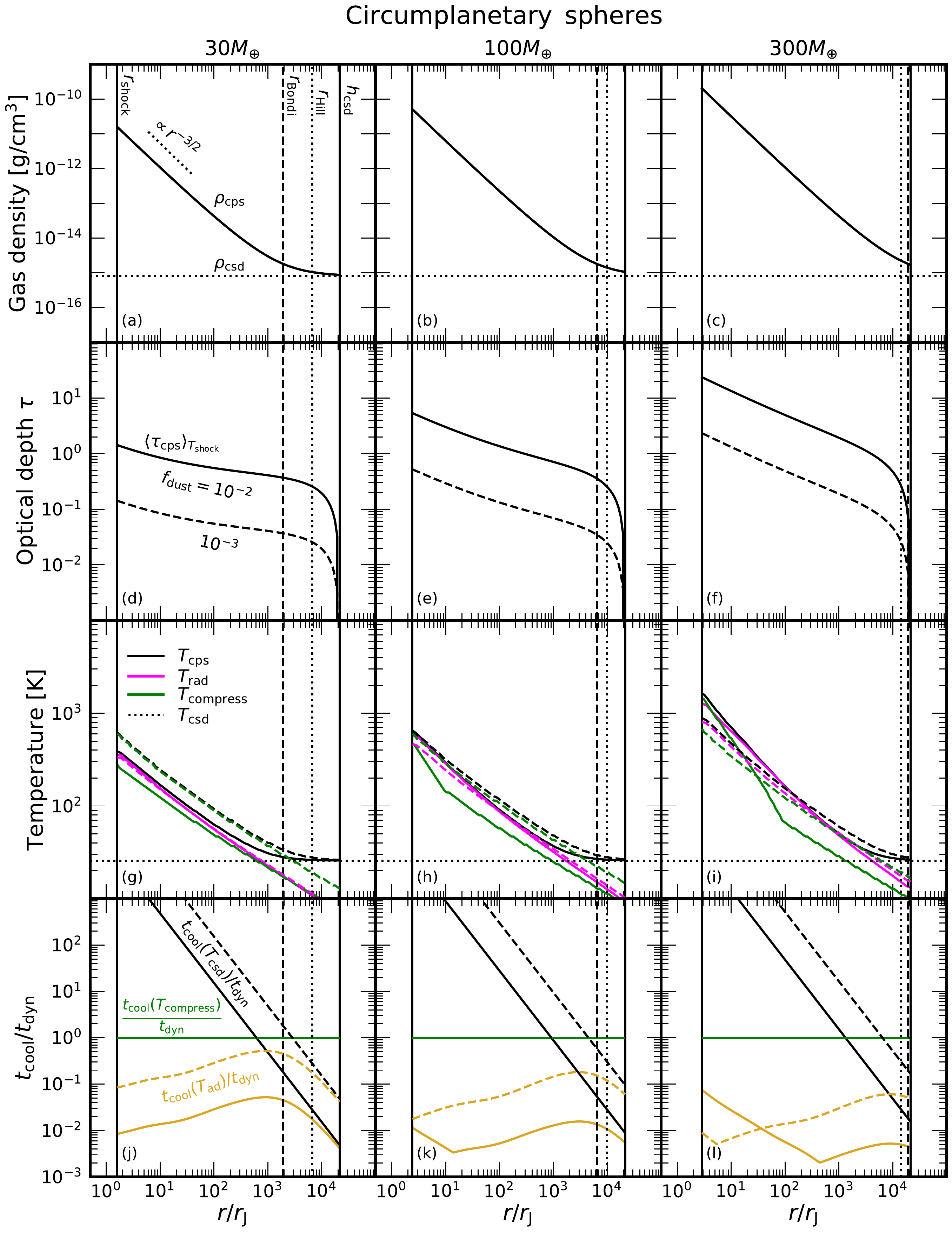}
\vspace{-0.45cm}
\caption{
CPS properties vs.~distance $r$ from the planet centre ($r_{\rm J}$ is Jupiter's radius).
Vertical solid lines running through each column mark the inner boundary of the CPS at the planetary accretion shock $r = r_{\rm shock}$, and the outer boundary at the CSD scale height $r = h_{\rm csd}$. Vertical dashed and dotted lines mark the planet's Bondi radius $r_{\rm Bondi} = Gm_{\rm p}/c_{\rm csd}^2$ and Hill radius $r_{\rm Hill} = (m_{\rm p}/\Mstar)^{1/3}a$, respectively. Our SED model is intended to describe planets in the ``subthermal'' regime where both of these scales are smaller than $h_{\rm csd}$. 
\textit{First row:} \referee{Gas density in the CPS, following the Bondi inflow profile with $\gamma = 7/5$.} When $r \ll r_{\rm Bondi}$, $\rho_{\rm cps} \propto r^{-3/2}$ (free fall). In the opposite limit, $\rho_{\rm cps}$ asymptotes to the background $\rho_{\rm csd}$ marked by the dotted horizontal line. The dust density (not shown) is lower than the gas density by a factor $f_{\rm dust} = \{10^{-3},\,10^{-2}\}$. \textit{Second row:} Radial optical depth from dust absorption, averaged over the Planck function at the planet's surface temperature $T_{\rm shock}$. In this and following rows, solid curves correspond to $f_{\rm dust} = 10^{-2}$ and dashed curves to $f_{\rm dust} = 10^{-3}$.
\referee{   \textit{Third row:} CPS temperatures.  Magenta curves show $T_{\rm rad}$, the temperature the CPS would have if only passively reprocessing radiation from $r_{\rm shock}$, calculated using $\texttt{radmc}$. Green curves show $T_{\rm compress}$, the temperature the CPS would have if heated only by compression during infall, calculated by setting the local radiative cooling time $t_{\rm cool}$ (equation \ref{eqn:tcool}) equal to the local infall time $t_{\rm dyn} = r /\sqrt{Gm_{\rm p}/r}$. Horizontal dotted lines mark the background $T_{\rm csd}$ set by stellar heating. The net CPS temperature is shown by the solid black curves and is defined by $T_{\rm cps}^4 = T_{\rm compress}^4 + T_{\rm rad}^4 + T_{\rm csd}^4$. 
\textit{Fourth row:} Demonstration that strictly isothermal and adiabatic CPS temperature profiles are not self-consistent. Gold curves plot $t_{\rm cool}/t_{\rm dyn}$ for the adiabatic CPS temperature $T_{\rm ad} = T_{\rm csd}(\rho_{\rm csd}/\rho_{\rm cps})^{\gamma-1}$. In this case $t_{\rm cool} \ll t_{\rm dyn}$, contradicting the adiabatic assumption. Black curves evaluate the cooling time if the CPS were isothermal at $T_{\rm csd}$ everywhere; then $t_{\rm cool} \gg t_{\rm dyn}$ across much of the envelope, contradicting the isothermal assumption. Our model assumes compression heats the CPS up to the point that $t_{\rm cool}/ t_{\rm dyn} = 1$ (green horizontal line), yielding $T_{\rm compress}$. Since $t_{\rm cool} \propto 1/T^3$, $T_{\rm compress}$ is a stable equilibrium.}
} 
  \label{fig:cps_summary}
\end{figure*}

\begin{figure*} 
\includegraphics[width=\textwidth]{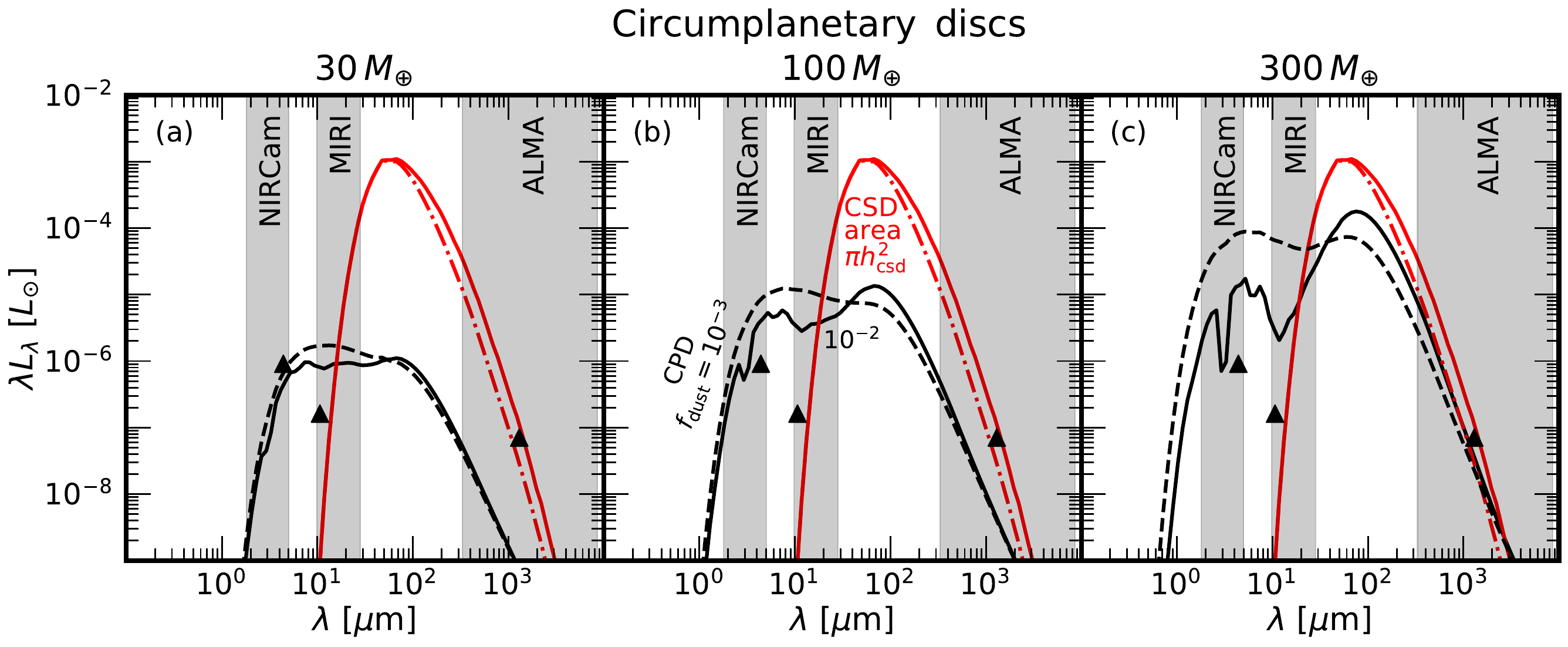} 
\caption{SEDs of protoplanets surrounded by circumplanetary discs (CPDs). Different panels correspond to different mass planets as labeled. Our modeled CPDs are heated by viscous dissipation in their interiors and by planetary radiation incident on their surfaces. The SED is not sensitive to the CPD viscosity parameter $\acpd$ (here fixed to $10^{-3}$) because the interior is optically thick. The planet+CPD radiation is re-processed by an outer CPS dust shell, assuming $f_{\rm dust} = 10^{-2}$ (black solid curve) or $f_{\rm dust} = 10^{-3}$ (black dashed curve). The solid red curve shows the SED of re-processed starlight from the local circumstellar disc (CSD), scaled to an emitting area $\pi h_{\rm csd}^2$ and plotted for solar $f_{\rm dust} = 10^{-2}$. Most of the CSD emission is generated by the surface layer (red dash-dot curve) and is insensitive to $f_{\rm dust}$. Shaded regions mark the passbands for coronographic imaging with JWST NIRCam and MIRI, and the wavelength range of ALMA. As in the spherically symmetric case (Fig.~\ref{fig:cps_sed}), protoplanets with CPDs seem most easily detectable against the local CSD with NIRCam and the shorter wavelengths of MIRI. \nick{The upward pointing triangles in the NIRCam and MIRI passbands mark detection limits for coronographic imaging without a background CSD in the F444W and F1065C bands, respectively, at a planet-star separation of 1 arcsecond (5$\sigma$ limit for F444W reported in fig. 5 of \citealt{carter_etal_2023}, and 3$\sigma$ limit for F1065C in fig. 5 of \citealt{boccaletti_etal_2022}; the contrast with the host star is 12.5 and 11.7 magnitudes, respectively).}
The triangle in the ALMA passband corresponds to a 1.3 mm continuum flux of 110 $\mu$Jy at a distance of 100 pc. This is the minimum flux needed for a point source to be detected against the background CSD according to injection tests performed on DSHARP gaps
(table 4 of \citealt{andrews_etal_2021}  
for a 90\% recovery fraction). }
  \label{fig:cpd_sed}
\end{figure*}

\begin{figure*} 
\includegraphics[width=\textwidth]{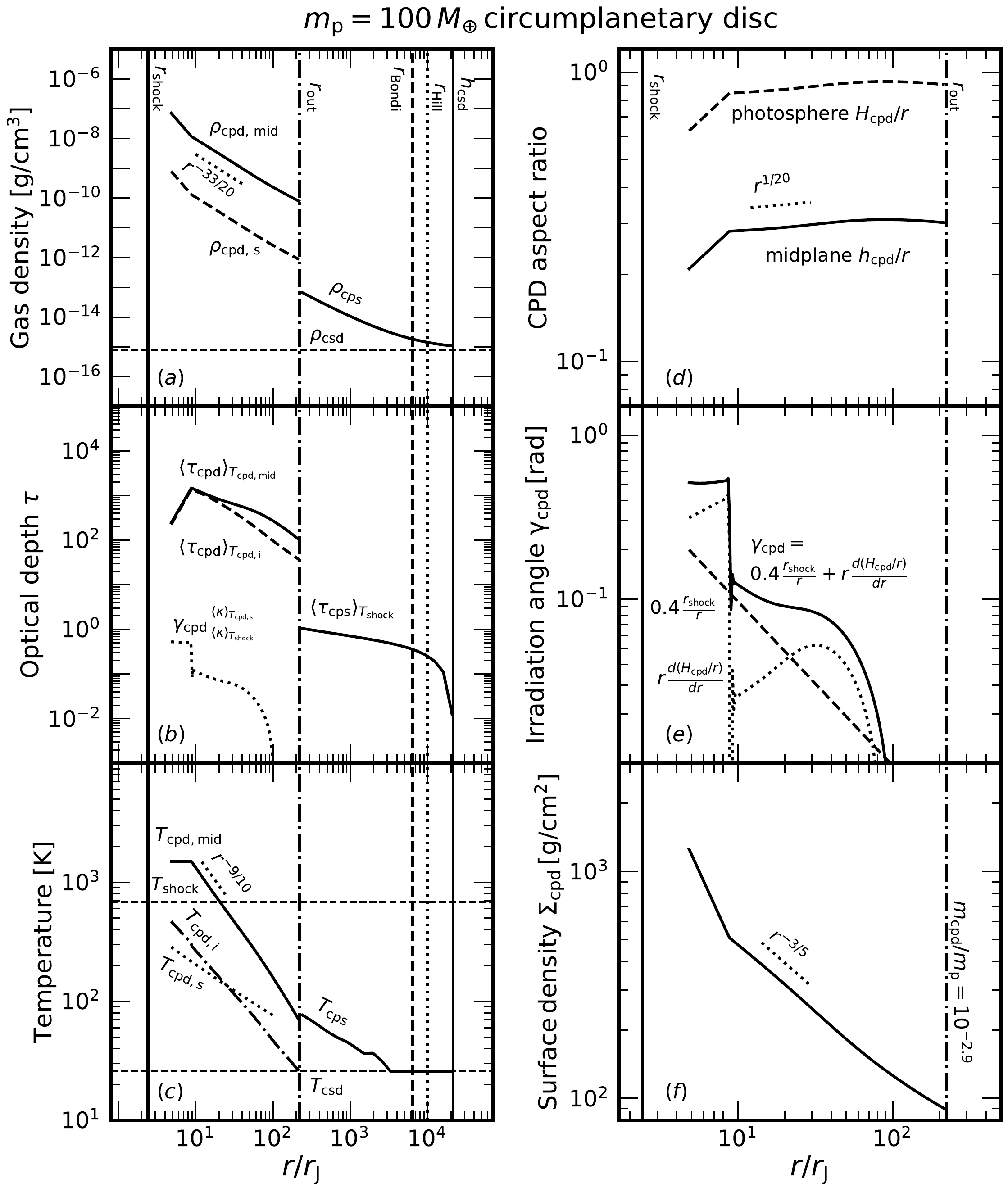}
\vspace{-0.7cm}
\caption{Our model $\acpd = 10^{-3}$ CPD around an $\Mp = 100M_\oplus$ planet accreting at $\dot{m}_{\rm p} = \Mp/t_{\rm age}$ (corresponding SED shown in the middle panel of Fig.~\ref{fig:cpd_sed}). The solid vertical line on the left of each panel marks the planetary accretion shock at $r = r_{\rm shock}$; the CPD inner edge is at $2r_{\rm shock}$ (boundary of a magnetospheric cavity). At $f_{\rm dust} = 10^{-2}$, the CPD is highly optically thick (panel b, solid curve) and midplane material is heated viscously to $T_{\rm cpd,mid}$ (panel c) assuming heat transport by radiative diffusion and a maximum temperature of 1500 K regulated by dust sublimation. The midplane temperature sets the CPD scale height $h_{\rm cpd}$ (panel d) and thus the midplane gas density (panel a) given the surface density (panel f), all solved for self-consistently  (eqs.~\ref{eqn:mdot_visc} and \ref{eqn:Tmid} and surrounding formulae). \referee{The disc surface layer at $H_{\rm cpd} = 3 h_{\rm cpd}$, with density $\rho_{\rm cpd,s} = \rho_{\rm cpd,mid}e^{-9/2}$, is heated to $T_{\rm cpd,s}$ by radiation from the central accretion shock and has a vertical optical depth (dotted line in panel b) determined by the angle $\gamma_{\rm cpd}$ at which this radiation grazes the disk surface (panel e). }The photosphere of the disc interior is heated to 
$T_{\rm cpd,i}$ by both irradiation and accretion. Where it falls below the background temperature $T_{\rm csd}$ defines the outer CPD radius $r_{\rm out}$ (dash-dot vertical line), beyond which we include a spherical dust shell (CPS), here marginally optically thin to planet light (for $f_{\rm dust} = 10^{-3}$, the optical depth would be ten times lower). The CPD (excluding CPS) mass is $m_{\rm cpd} \approx 0.1\,\Mearth$. Note the change in radial scale between left and right columns. 
}
  \label{fig:cpd_summary}
\end{figure*}


\begin{figure*} 
\vspace{-1.2cm}
\includegraphics[width=0.85\textwidth]{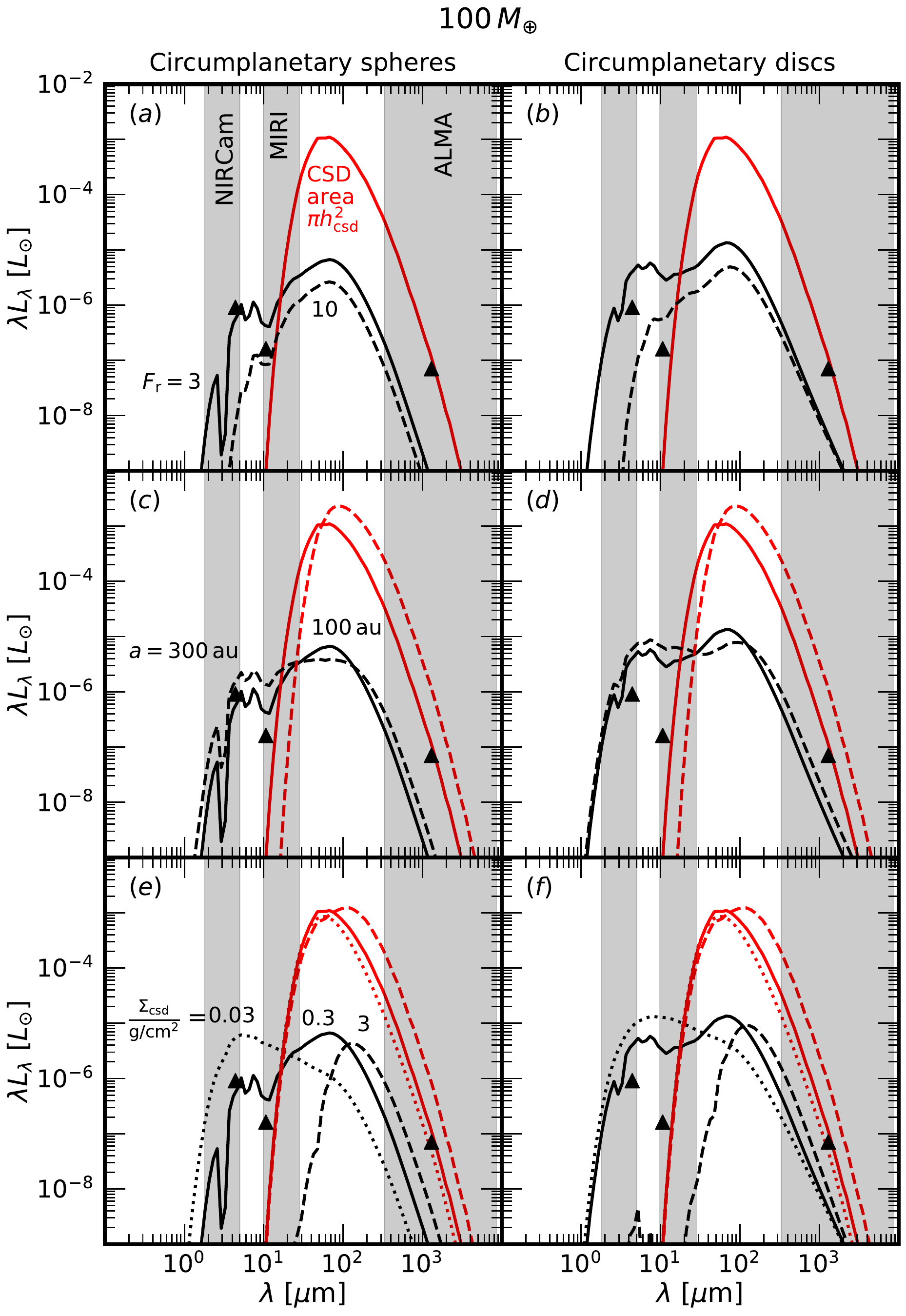}
\caption{How CPS SEDs (left column) and CPD SEDs (right column) depend on input parameters, at fixed $m_{\rm p} = 100\,\Mearth$ and solar dust-to-gas ratio $f_{\rm dust} = 10^{-2}$. In each column, the black solid curve plots the emergent protoplanet SED for a fiducial parameter set $\{F_{\rm r},\,a,\,\Sigma_{\rm csd}\} = \{3,\,100\,\mathrm{au},\,0.3\,\rm g/cm^2\}$ and is the same across all panels vertically. Each panel in a column varies one of these parameters at a time, with different linestyles corresponding to different parameter choices as labeled. \textit{Top row:} How SEDs depend on the protoplanet's normalised radius $F_{\rm r}$. More compact protoplanets (smaller $F_{\rm r}$) are brighter and have bluer SEDs less masked by CSD emission (red curve). \textit{Middle row:} How SEDs depend on the planet's orbital radius $a$. The background CSD temperature is scaled using $T_{\rm csd} \propto a^{-3/7}$ \protect \citep{chiang_goldreich_1997}. Increasing $a$ boosts protoplanet detectability at short wavelengths by shifting CSD emission to longer wavelengths. \textit{Bottom row:} How SEDs depend on the local surface density $\Sigma_{\rm csd}$ (inside whatever circumstellar gap the planet may have opened). Higher densities increase the optical thickness of the CPS (or outer circumplanetary shell in CPD models) and lead to stronger attenuation of the protoplanet's near-infrared accretion luminosity. The CSD SED changes by at most a factor of a few between the different $\Sigma_{\rm csd}$ values because for these values the CSD stays optically thick to starlight (contrast with PDS 70 as shown in Fig.~\ref{fig:pds70_sed}).
}
  \label{fig:params}
\end{figure*}

\begin{figure*} 

\includegraphics[width=\columnwidth]{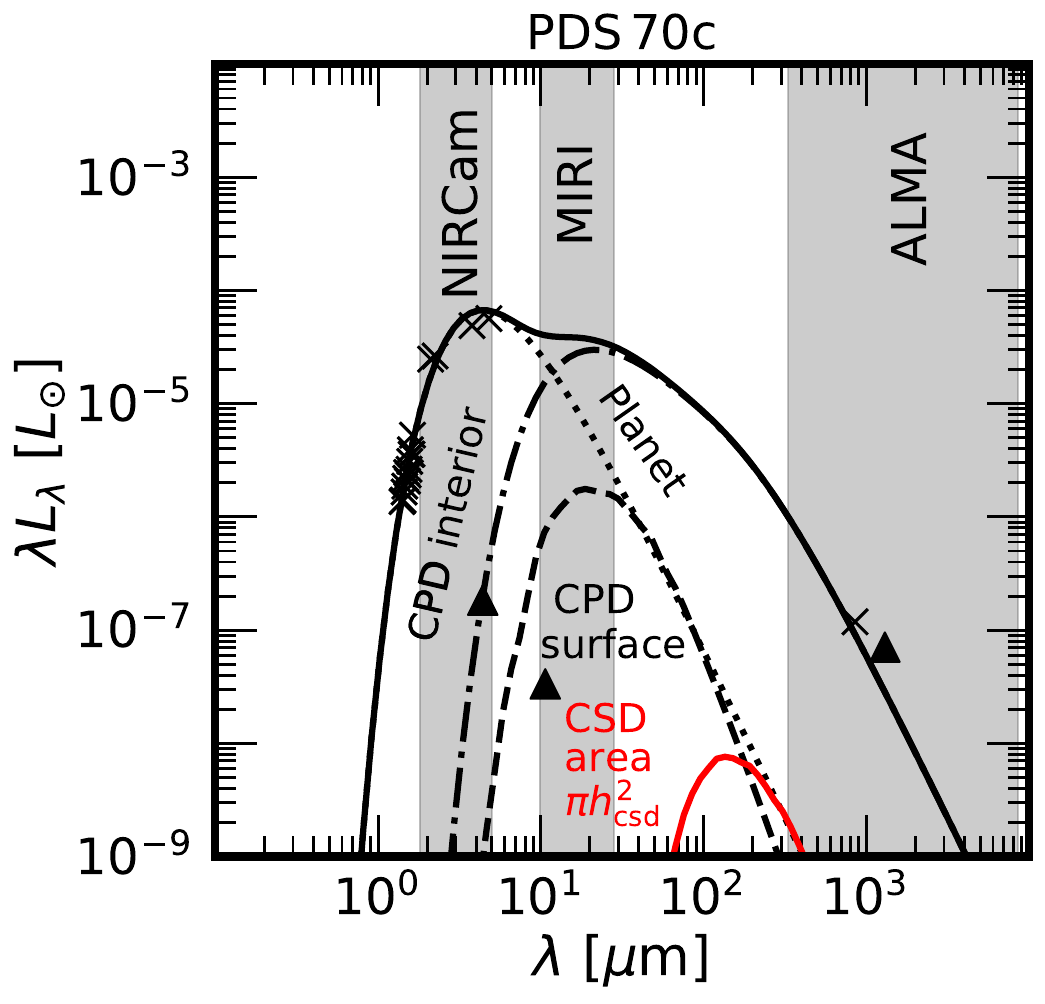}
\caption{Reproducing the observed SED of the protoplanet PDS 70c (crosses; \citealt{mesa_etal_2019, wang_etal_2020, wang_etal_2021, benisty_etal_2021, christiaens_etal_2024}) using a pure-CPD model for $\{m_{\rm p},\,t_{\rm age},\,f_{\rm dust},\,T_{\rm csd}\} = \{3\,m_{\rm J},\,5\,\mathrm{Myr},\,1.5\times 10^{-3},\,20\,\mathrm{K}\}$ (table 1 of \citealt{choksi_etal_2023}; \citealt{portilla-revelo_etal_2023}), an accretion shock radius $\rshock = 4.5\,r_{\rm J}$ (cf.~\citealt{wang_etal_2020}), a CPD inner edge $r_{\rm in} = 5r_{\rm shock}$ (possibly the result of magnetospheric truncation), and an accretion rate $\dot{m}_{\rm p} = m_{\rm p}/t_{\rm age}$. The total model SED (solid curve) is the sum of the emission from the planet's accretion shock (dotted curve) and the CPD's interior (dash-dotted) and surface (dashed). Our model CPDs are largely self-luminous, with viscous heating dominating passive reprocessing (see section \ref{subsec:cpd_results}). Unlike our standard CPD models, our model for PDS 70c omits the surrounding CPS envelope on Bondi scales because $\rb > h_{\rm csd}$. \referee{PDS 70c sits inside a highly depleted transitional disc cavity having a gas surface density $\Sigma_{\rm csd} = 8 \times 10^{-3}\,\gcm$, as inferred from observations of CO isotopologues \citep{choksi_chiang_2022}. This surface density is a factor of 40 lower than in our fiducial models, and renders the PDS 70 cavity optically thin to incident starlight; the local CSD SED is modeled using a single, optically thin blackbody having $(\lambda L_{\lambda})_{\rm csd} = 4\pi^2 h_{\rm csd}^2\lambda B_{\lambda}(T_{\rm csd})\kappa(\lambda)\Sigma_{\rm csd}f_{\rm dust}$. From a combination of multiple factors, the local CSD is orders of magnitude fainter than in our fiducial models (see section \ref{subsec:pds_results} for details).} 
The detection limits in the NIRCam and MIRI passbands correspond to the same contrasts with the host star as in previous figures, but now reflect a stellar radius and temperature of $R_{\star} = 1.1\,R_{\odot}$ and $T_{\star} = 4100$ K  \protect \citep{keppler_etal_2019}.}
\label{fig:pds70_sed}
\end{figure*}

\begin{figure*} 
\includegraphics[width=\textwidth]{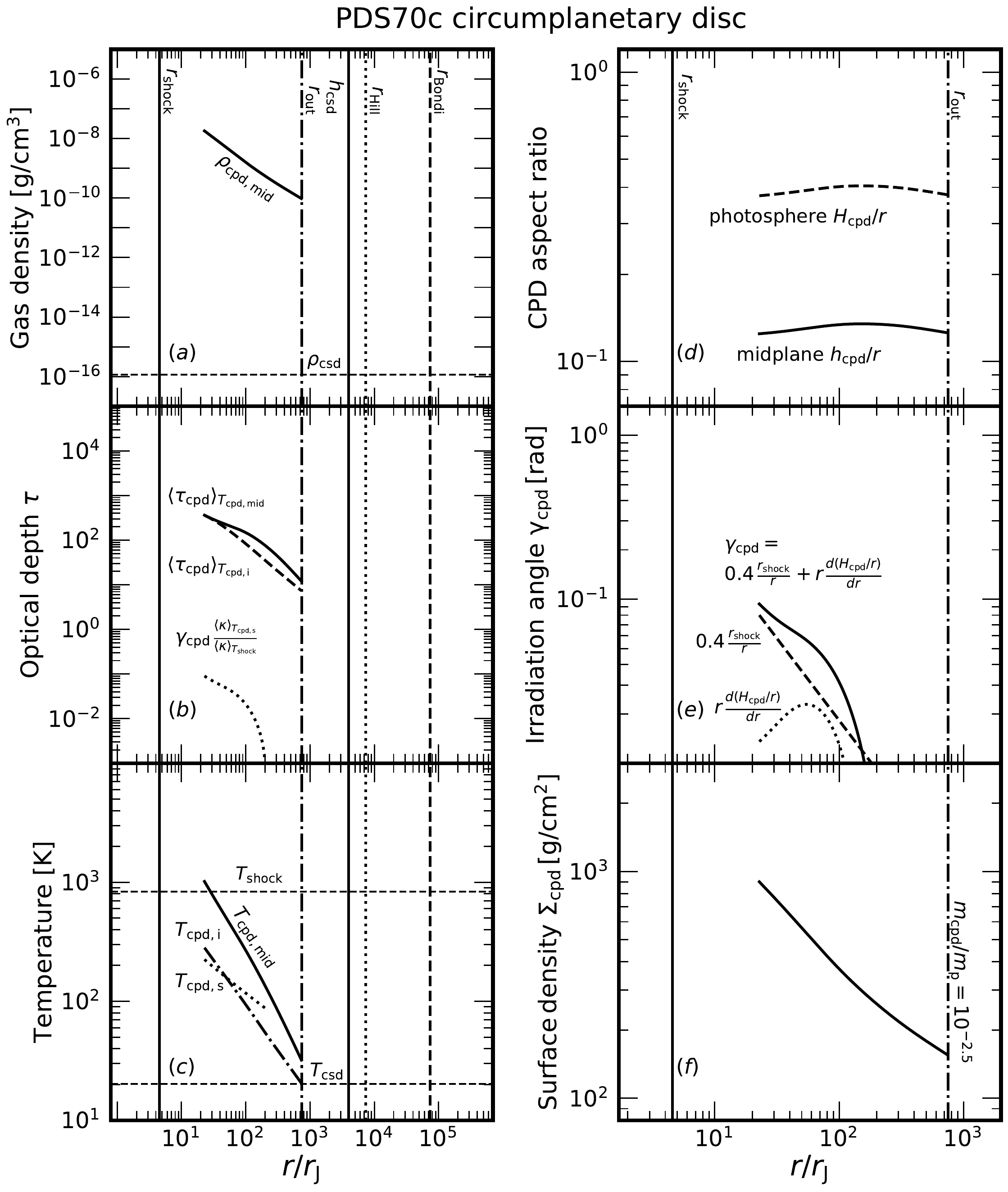}
\vspace{-7mm}
\caption{ 
Diagnostics of a pure-CPD model for PDS 70c that reproduces its observed SED (see the caption to Fig.~\ref{fig:pds70_sed} for parameter choices, and the caption to Fig.~\ref{fig:cpd_summary} for descriptions of the variables plotted). For PDS 70c, $r_{\rm Bondi} > r_{\rm Hill} > h_{\rm csd}$, and so we omit the CPS envelope on Bondi scales that surrounds the CPD in our standard models; here the model ends at the CPD outer radius $r_{\rm out}$ (where $T_{\rm cpd,i}$ falls below $T_{\rm csd}$, panel c). The outermost portion of the CPD is shadowed from the host planet (see panels d and e where $d(H_{\rm cpd}/r)/dr$ and the irradiation grazing angle $\gamma_{\rm cpd}$ drop below zero). In these regions, passive reprocessing, which was dominated by viscous heating even in unshadowed regions, is omitted from the energy budget. Note the change in radial scale between left and right columns. 
}
\label{fig:pds70_summary}
\end{figure*}

\section{Results}
\label{sec:results}
We describe SEDs of pure circumplanetary sphere (CPS) models in section \ref{subsec:cps_results}, and combined circumplanetary disc (CPD) + sphere models in section \ref{subsec:cpd_results}. In those sections we concentrate on the effects of varying planet mass $m_{\rm p}$ and dust-to-gas ratio $f_{\rm dust}$; other parameters of protoplanet radius $F_r$, stellocentric radius $a$, and circumstellar disc surface density $\Sigma_{\rm csd}$ are explored in section \ref{subsec:params}. An application of our model to the protoplanet PDS 70c is presented in section \ref{subsec:pds_results}.

\subsection{SEDs of circumplanetary spheres (CPSs)} \label{subsec:cps_results}

\referee{Figure \ref{fig:cps_sed} shows SEDs for accreting protoplanets surrounded by spherically symmetric envelopes having different dust-to-gas ratios (black curves). For $f_{\rm dust} = 10^{-3}$ (dashed black), the CPSs are largely optically thin in the near-infrared (marginally so for $m_{\rm p} = 300\,m_{\oplus}$; Figure \ref{fig:cps_summary}d-f). The emergent power at these wavelengths traces the underlying blackbody emission from the accretion shock: for protoplanets that are $F_{\rm r} = 3$ times puffier than mature (cooled) planets, the bulk of the accretion power is radiated at wavelengths 3-6 $\mu$m, corresponding to temperatures $T_{\rm shock} \approx 500$-1000 K.} 


\referee{Solar metallicity CPSs ($f_{\rm dust} = 10^{-2}$) can be orders of magnitude fainter in the near-IR (solid black curves in Fig.~\ref{fig:cps_sed}) because they obscure the radiation from the planet, with optical depths ranging from $\sim$2-20 (Fig. \ref{fig:cps_summary}d-f). The photospheres of these CPSs are cold enough to retain water ice on grains ($T < 150$ K; Fig.~\ref{fig:cps_summary}g-i); the 2.7 $\mu$m vibrational resonance from ice appears in absorption and carves out even more near-infrared flux. Surprisingly, Figure \ref{fig:cps_sed} also shows that for $f_{\rm dust} = 10^{-2}$, more massive protoplanets are fainter in the near-IR. This trend is a consequence of more massive planets pulling down greater amounts of gas and dust into their Bondi spheres, so that their envelopes are optically thicker (Fig. \ref{fig:cps_summary}a-c). The extinction $e^{-\tau}$ increases with planet mass more steeply than does the accretion luminosity.}


\referee{Power at wavelengths 10-1000 $\mu$m comes from heated CPS dust. In our modeled CPSs, compressional and radiative heating are competitive with each other (cf. \citealt{adams_batygin_2022} and \citealt{ taylor_adams_2024} who neglect compression). Fig. \ref{fig:cps_summary}g-i demonstrates this by plotting the temperatures $T_{\rm rad}$ and $T_{\rm compress}$ that the CPS would have if each heat source were considered in isolation, and showing that the two temperatures are comparable, with $T_{\rm compress} > T_{\rm rad}$ for $f_{\rm dust} = 10^{-3}$ and $m_{\rm p} \lesssim 100\,m_{\oplus}$.}

Fig. \ref{fig:cps_sed} shows that for $\lambda \gtrsim 20 \, \mu$m, the local circumstellar disc (CSD, red curve) is orders of magnitude brighter than the CPS. Prospects for directly imaging an embedded planet are much better at $\lambda \lesssim 10 \,\mu$m where contamination from the CSD is weaker. Fig.~\ref{fig:cps_sed} plots  detection limits for near-infrared coronographic imaging with JWST's NIRCam (the James Webb Space Telescope's near-infrared camera) and MIRI (mid-infrared instrument).  
At $\lambda \approx 4 \,\mu$m, protoplanets more massive than Saturn ($\sim$$100 \, \Mearth$) may be bright enough to be detected with NIRCam against the glare of their host stars, if the dust-to-gas ratio in the circumplanetary environment is sufficiently sub-solar. A similar statement applies at $\lambda \approx 11 \,\mu$m for $m_{\rm p} \gtrsim 30 \, m_{\oplus}$ imaged with MIRI.

\subsection{SEDs of circumplanetary discs (CPDs)}
\label{subsec:cpd_results}

Figure \ref{fig:cpd_sed} shows SEDs for hybrid models in which the protoplanet accretes from a circumplanetary disc (CPD) surrounded by a CPS shell, for dust-to-gas ratios of $f_{\rm dust} = 10^{-3}$ (dashed curves) and $10^{-2}$ (black curves). Near-infrared emission originates from the planetary accretion shock, mid-infrared emission from the CPD, and far-infrared and longer wavelength radiation from the CPS shell. As $f_{\rm dust}$ increases, the CPS shell obscures and reprocesses more of the protoplanet's emission.




Most of the CPD emission comes from the viscously heated interior and not from the passively re-processing surface layer. Viscous dissipation dominates because our CPDs flare only weakly ($H_{\rm cpd}/r \propto h_{\rm cpd}/r \propto r^{1/20}$; Fig.~\ref{fig:cpd_summary}d), keeping the radiation grazing angle $\gamma_{\rm cpd}$ low (Fig.~\ref{fig:cpd_summary}e), and by extension the optical depth of the surface layers low (Fig.~\ref{fig:cpd_summary}b, dotted curve).\footnote{The scalings of effective temperature with radius are identical ($T_{\rm eff} \propto r^{-3/4}$) between a viscously heated disc accreting at some steady rate, and a perfectly flat disc passively reprocessing light from a central source powered by accretion at the same rate (e.g.~\citealt{shu_1992}). But the order-unity coefficients are such that the emitted fluxes ($\propto T_{\rm eff}^4$) of the former exceed those of the latter by about a factor of 10.}


As in spherically symmetric pure-CPS models, the emergent SED from CPD models is masked by circumstellar (CSD) emission longward of 20 $\mu$m (red curve in Fig.~\ref{fig:cpd_sed}). We conclude that, regardless of how circumplanetary material is spatially distributed, detection prospects for embedded planets are best at $\lambda \lesssim 10$ $\mu$m, where emission from the relatively hot accretion shock (plus radiation from the planet's interior, not included in our SEDs) stands out against the cooler circumstellar disc. The planetary near-infrared emission is less extincted when circumplanetary material is flattened into a face-on disc; Saturn and higher-mass planets can conceivably be detected with JWST NIRCam in our CPD models (Fig.~\ref{fig:cpd_sed}) but may not be in our CPS models (Fig.~\ref{fig:cps_sed}). At $\lambda \approx 11$ $\mu$m, detection of 30 $m_\oplus$ planets with MIRI appears possible in both CPD and CPS models.

\referee{In our simplistic hybrid model, the gas density is not continuous at the CPD/CPS boundary  (Fig.~\ref{fig:cpd_summary}a). A more physical model would better capture density gradients there. However, whatever artefacts our density discontinuity induces in the protoplanet SED are likely to be masked by the local CSD SED, since at the CPD/CPS boundary, the CPD interior temperature grades into the CSD temperature, by construction.}

 



\subsection{Other model parameters}
\label{subsec:params}



Figure \ref{fig:params} explores how CPS and CPD SEDs depend on the normalised protoplanet size $F_{\rm r}$, orbital radius $a$, and background gas surface density $\Sigma_{\rm csd}$. We vary these parameters one at a time while holding the others fixed at $\{F_{\rm r},\,a,\,\Sigma_{\rm csd}\} = \{3,\,100\,\mathrm{au},\,0.1\,\rm g/cm^2\}$, the values used in Figs. \ref{fig:cps_sed}-\ref{fig:cpd_summary}.

Puffier protoplanets (larger $F_{\rm r}$) are more difficult to detect because they have lower accretion luminosities (at fixed $\Mp$ and $\dot{m}_{\rm p}$) and therefore less emission overall (Fig.~\ref{fig:params}a-b). 
Increasing the planet's orbital radius $a$ does not much affect the emergent SED in either CPS or CPD models (Fig.~\ref{fig:params}c-d). However, farther from the star, the CSD is cooler (at $a = 300$ au we set $T_{\rm csd,s} = 30$ K)  and so its SED shifts to longer wavelengths, opening up the possibility of detecting protoplanet emission against the CSD background with MIRI (JWST's mid-infrared instrument). 
A lower background surface density $\Sigma_{\rm csd}$ reduces the extinction from the CPS (or outer CPS shell in CPD models), allowing more of the near-infrared flux from the accreting protoplanet to emerge unimpeded (Fig.~\ref{fig:params}e-f).

We found that changing the disc viscosity parameter $\acpd$ from $10^{-4}$ to $10^{-2}$ does not noticeably affect the SED (data not shown). Increasing $\acpd$ reduces the mass in the CPD (at assumed fixed $\dot{m}_{\rm p})$, but the CPD emission is independent of mass insofar as it comes mostly from the photosphere of the optically thick interior.

\subsection{Application to PDS 70c}
\label{subsec:pds_results}
We apply our model to reproducing the SED of PDS 70c, a confirmed protoplanet of mass $m_{\rm p} \sim 1$-$10\, m_{\rm J}$ and age $t_{\rm age} \sim 5$ Myr (for these and other parameters of the PDS 70 system, see \citealt{choksi_etal_2023}, \citealt{choksi_chiang_2022}, and references therein). The planet is located 34 au from a 
$M_{\star} = 0.88\,\Msun$ star within a transitional circumstellar disc cavity of surface density $\Sigma_{\rm csd} \sim 0.008$ g/cm$^2$, dust-to-gas ratio $f_{\rm dust} \approx 1.5 \times 10^{-3}$, and midplane temperature $T_{\rm csd} \approx 20$ K (\citealt{portilla-revelo_etal_2023}). Given the above parameters, PDS 70c is ``superthermal'', i.e.~its gravitational sphere of influence is limited not to its Bondi sphere but to its Hill sphere, whose radius exceeds the CSD gas scale height. 
Accordingly, most of the CSD material may accrete into the Hill sphere at lower planet latitudes, closer to the disc midplane (see e.g.~figure 7 of \citealt{choksi_etal_2023}). We are therefore motivated to consider a pure-CPD model for PDS 70c, omitting the surrounding CPS shell on Bondi scales. We do not model PDS 70b, but note that its mass and observed near-infrared and mm-wave brightness are similar to those of PDS 70c (see, e.g., \citealt{choksi_chiang_2022}).

\proofs{Figure \ref{fig:pds70_sed} shows our model SED (solid curve) fitted by eye to the observations (crosses). All the near-infrared flux from $\lambda \approx 1$ to 4 $\mu$m originates from the accretion shock, a spherical blackbody of mass $m_{\rm p} = 3 \, m_{\rm J}$ and radius $r_{\rm shock} = 4.5 \, r_{\rm J}$, accreting at rate $\dot{m}_{\rm p} = m_{\rm p}/t_{\rm age} = 0.6\,m_{\rm J}/\mathrm{Myr}$. Our adopted accretion rate $\dot{m}_{\rm p}$ is within a factor of 2 of the theoretical maximum of $\dot{m}_{\rm in} \approx 1\,m_{\rm J}/\mathrm{Myr}$ computed according to equation (18) of \citet{choksi_etal_2023} for our parameters. Our fitted value of $r_{\rm shock}$ corresponds to $F_r \approx 4.5$, within the range expected from protoplanet cooling models (see discussion following equation \ref{eqn:rshock}). }

\proofs{The radius of the CPD inner edge is fitted to be $r_{\rm in} \gtrsim 5r_{\rm shock} \approx 22.5 r_{\rm J}$. Smaller values of $r_{\rm in}$ result in the disc contributing too much flux in the near-infrared. Larger values may be constrained with future long-wavelength MIRI observations. If $r_{\rm in}$ is set by the planet's magnetosphere, the magnetic field strength at the planet's effective surface at $r_{\rm shock}$ is $B \gtrsim 200$ Gauss, assuming a dipole field \citep[e.g.][]{ghosh_lamb_1998}. This constraint is consistent with expectations from a convective dynamo which predicts $B \sim (m_{\rm p}/r_{\rm shock}^3)^{1/6}(L_{\rm acc}/r_{\rm shock}^2)^{1/3} \sim 500$ Gauss \citep{christensen_etal_2009}.}

\proofs{The mm-wave flux measured by the Atacama Large Millimeter Array (ALMA; \citealt{benisty_etal_2021}) originates from the cool outer regions of the CPD. The disc extends to $r_{\rm out} \approx 700r_{\rm J} \approx r_{\rm Hill}/10$, small enough that it avoids being resolved by ALMA \citep{benisty_etal_2021}. For our assumed $\alpha_{\rm cpd} = 10^{-3}$, the CPD mass is $m_{\rm p}/300$. For further model details, see Figure \ref{fig:pds70_summary}. }


\referee{Fig. \ref{fig:pds70_sed} also shows that the local CSD is orders of magnitude fainter than the CPD, unlike in typical annular ALMA gaps (contrast with Figs.~\ref{fig:cps_sed} and \ref{fig:cpd_sed}). 
The difference arises from a number of factors, one of which is that the transitional disc cavity hosting the PDS 70 protoplanets is optically thin even to host star radiation, and is therefore modeled as a single dilute blackbody with $(\lambda L_{\lambda})_{\rm csd} = 4\pi^2 h_{\rm csd}^2 \lambda B_{\lambda}(T_{\rm csd})\tau(\lambda)$ (cf. equation \ref{eqn:csd_sed}). The optical depth at the blackbody's peak wavelength is $\tau \approx 4 \times 10^{-4}$, a factor of 100 smaller than the maximum vertical (perpendicular to the midplane) optical depth of $0.25 \times h_{\rm csd}/a \approx 0.03$ for the fiducial CSD's surface layer (the factor of 0.25 is the ratio of emissivities at the peak emitting and absorbing wavelengths). Second, the CSD temperature of $T_{\rm csd} = 20$ K from mm-wave observations is lower than the 50 K surface layer temperature adopted in our fiducial models and accounts for another factor of $\sim$$(50/20)^4 = 40$ reduction in luminosity. Finally, the area $\pi h_{\rm csd}^2$ of the emitting patch is a factor of 36 smaller at the location of PDS 70c compared to our fiducial model.}

\begin{table*}
\begin{adjustwidth}{-1.73cm}{}
\scalebox{0.95}{
\begin{tabular}{cccccccccccccc}
\hline 
(1) & (2) & (3) & (4) & (5) & (6) & (7) & (8) & (9) & \nickk{(10)} & (11) & (12) & (13) & (14) \\ 
Name & $d$ & $M_{\star}$ &  $t_{\rm age}$ & $a$ & $\Sigma_{\rm csd}$ & $T_{\rm csd}$ & $m_{\rm p}$ &  $\dot{m}_{\rm p}$ & \nickk{$r_{\rm shock}$} & $L_{\rm acc}$ & $T_{\rm shock}$ & $\lambda_{\rm peak}$ & $F_{\nu}$ \\

& $[\rm pc]$  & $[M_{\odot}]$ &   $[\rm Myr]$ & $[\rm au]$ & $[\rm g/cm^2]$ & $[\rm K]$ &  $[m_{\rm J}]$ &$[m_{\rm J}/ \rm Myr]$ & \nickk{$[r_{\rm J}]$} & $[L_{\odot}]$ & $[\rm K]$ &  $[\mu \rm m]$ & $[\mu \rm Jy]$ \\
\hline
Sz 114 & $162$ & 0.17 & $1$ & 39 & ... & $10$ & 0.01-0.02 & 0.01-0.02 & 0.8-1 & $10^{-7.4}\mathrm{-}10^{-6.9}$ & 290-340 & 10-20 & 0.2-0.4 \\
GW Lup & $155$ & 0.46 & $2$ & 74 & ... & $10$ & 0.007-0.03 & 0.003-0.02 & 0.7-1 & $10^{-8}\mathrm{-}10^{-7}$ & 200-300 & 15-23 & 0.06-0.6 \\
Elias 20 & $138$ & 0.48 & $0.8$ & 25 & ... & $30$ & 0.03-0.07 & 0.03-0.08 & 1.1-1.4 & $10^{-7}\mathrm{-}10^{-6}$ & 400-500 & 10-13 & 1-4 \\
Elias 27 & $116$ & 0.49 & $0.8$ & 69 & ... & $10$ & 0.01-0.07 & 0.01-0.08 & 0.8-1 & $10^{-7}\mathrm{-}10^{-6}$ & 300-500 & 10-20 & 0.4-6 \\
RU Lup & $159$ & 0.63 & $0.5$ & 29 & ... & $30$ & 0.03-0.07 & 0.07-0.1 & 1.1-1.4 & $10^{-6.3}\mathrm{-}10^{-5.8}$ & 500-600 & 9-10 & 2-4 \\
SR 4 & $134$ & 0.68 & $0.8$ & 11 & ... & $40$ & 0.2-2 & 0.2-3 & 2-3 & $10^{-5}\mathrm{-}10^{-3}$ & 600-2000 & 3-8 & 20-700 \\
Elias 24 & $136$ & 0.8 & $2$ & 55 & ... & $30$ & 0.5-5 & 0.2-2 & 2.6-3.0 & $10^{-5}\mathrm{-}10^{-3}$ & 700-2000 & 3-7 & 40-1000 \\
TW Hya-G1 & $60$ & 0.8 & $8$ & 21 & 0.04-3 & $60$ & 0.03-0.3 & 0.003-0.03 & 1-2 & $10^{-8}\mathrm{-}10^{-6}$ & 200-400 & 10-20 & 1-30 \\
TW Hya-G2 & $60$ & 0.8 & $8$ & 85 & 0.008-0.2 & $20$ & 0.02-0.2 & 0.002-0.02 & 0.9-2 & $10^{-8}\mathrm{-}10^{-6}$ & 200-300 & 20-30 & 0.5-10 \\
Sz 129 & $161$ & 0.83 & $4$ & 41 & ... & $20$ & 0.02-0.03 & 0.004-0.008 & 0.9-1 & $10^{-8}\mathrm{-}10^{-7}$ & 200-300 & 19-22 & 0.1-0.3 \\
DoAr 25-G1 & $138$ & 0.95 & $2$ & 98 & ... & $10$ & 0.07-0.1 & 0.03-0.05 & 1-2 & $10^{-6.4}\mathrm{-}10^{-6.1}$ & 390-430 & 12-13 & 2-4 \\
DoAr 25-G2 & $138$ & 0.95 & $2$ & 125 & ... & $10$ & 0.02-0.03 & 0.008-0.02 & 0.9-1 & $10^{-7.4}\mathrm{-}10^{-6.9}$ & 280-330 & 15-19 & 0.3-0.7 \\
IM Lup & $158$ & 1.1 & $0.5$ & 117 & 0.1-10 & $20$ & 0.03-0.1 & 0.07-0.2 & 1-2 & $10^{-6}\mathrm{-}10^{-5}$ & 500-600 & 8-10 & 2-8 \\
AS 209-G1 & $121$ & 1.2 & $1$ & 9 & ... & $50$ & 0.2-2 & 0.2-2 & 2-3 & $10^{-5}\mathrm{-}10^{-3}$ & 600-2000 & 3-8 & 20-700 \\
AS 209-G2 & $121$ & 1.2 & $1$ & 99 & 0.04-0.4 & $10$ & 0.1-0.7 & 0.1-0.7 & 2-3 & $10^{-6}\mathrm{-}10^{-4}$ & 500-900 & 6-10 & 8-100 \\
$\rm AS\,209{\text -}G3$ & $121$ & 1.2 & $1$ & 240 & 0.0002-0.1 & $7$ & 0.01-0.05 & 0.01-0.05 & 0.8-1 & $10^{-7}\mathrm{-}10^{-6}$ & 300-400 & 10-20 & 0.4-3 \\
HD 142666 & $148$ & 1.58 & $10$ & 16 & ... & $60$ & 0.03-0.3 & 0.003-0.03 & 1-2 & $10^{-8}\mathrm{-}10^{-6}$ & 200-400 & 10-20 & 0.2-5 \\
HD 169142 & $115$ & 1.65 & $10$ & 37 & 0.1-0.2 & $60$ & 0.1-1 & 0.01-0.1 & 2-3 & $10^{-7}\mathrm{-}10^{-5}$ & 300-600 & 8-20 & 2-40 \\
$\mathrm{MWC\,758}$ & $160$ & 1.9 & $10$ & 100 & ... & $200$ & 3-100 & 0.3-10 & 3-2 & $10^{-4}\mathrm{-}10^{0}$ & 1000-9000 & 0.6-5 & 100-20000 \\
HD 143006-G1 & $165$ & 1.78 & $4$ & 22 & ... & $30$ & 1-20 & 0.3-5 & 2.9-2.7 & $10^{-4}\mathrm{-}10^{-2}$ & 900-4000 & 1-6 & 60-4000 \\
HD 163296-G1 & $101$ & 2.0 & $10$ & 10 & ... & $300$ & 0.1-0.7 & 0.01-0.07 & 2-3 & $10^{-7}\mathrm{-}10^{-5}$ & 300-500 & 10-20 & 2-30 \\
HD 163296-G2 & $101$ & 2.0 & $10$ & 48 & 1-40 & $70$ & 0.3-2 & 0.03-0.2 & 2-3 & $10^{-6}\mathrm{-}10^{-4}$ & 400-900 & 6-10 & 9-200 \\
HD 163296-G3 & $101$ & 2.0 & $10$ & 86 & 0.1-20 & $40$ & 0.03-1 & 0.003-0.1 & 1-3 & $10^{-8}\mathrm{-}10^{-5}$ & 200-600 & 8-20 & 0.4-50 \\
$\rm HD\,163296{\text -}G4$ & $101$ & 2.0 & $10$ & 137 & 0.2-7 & $30$ & 0.002-1 & 0.0002-0.1 & 0.5-3 & $10^{-9}\mathrm{-}10^{-5}$ & 100-700 & 8-40 & 0.01-80 \\
$\rm HD\,163296{\text -}G234alt$ & $101$ & 2.0 & $10$ & 108 & 0.5-10 & $30$ & 0.2 & 0.02 & 2.0 & -6.0 & 300.0 & 10.0 & 6.0 \\
$\rm HD\,163296{\text -}G5$ & $101$ & 2.0 & $10$ & 260 & 0.1-2 & $20$ & 0.01-2 & 0.001-0.2 & 0.8-3 & $10^{-8}\mathrm{-}10^{-4}$ & 200-800 & 6-30 & 0.1-200 \\
AB Aur & $163$ & 2.4 & $3$ & 94 & ... & $30$ & 9-100 & 3-40 & 3-2 & $10^{-3}\mathrm{-}10^{0}$ & 2000-10000 & 0.3-2 & 1000-90000 \\
PDS 70b & $113$ & 0.88 & $5$ & 22 & 0.0008-0.08 & $30$ & 1-10 & 0.2-2 & 2.88-2.91 & $10^{-5}\mathrm{-}10^{-3}$ & 700-2000 & 2-7 & 70-2000 \\
PDS 70c & $113$ & 0.88 & $5$ & 34 & 0.0008-0.08 & $20$ & 1-10 & 0.2-2 & 2.88-2.91 & $10^{-5}\mathrm{-}10^{-3}$ & 700-2000 & 2-7 & 70-2000 \\ \hline 
\end{tabular}
}
 \end{adjustwidth}
\caption{Properties of gapped discs and the planets hypothesized to reside within them (confirmed in the case of PDS 70), adapted from \protect\cite{choksi_etal_2023}. 
Column headings: (1) System name. We append ``G\#'' to distinguish between different gaps in a given system. \nick{The MWC 758 and AB Aur entries were not tabulated by \protect \cite{choksi_etal_2023} and are further documented in appendix \ref{appendix:data}.} (2) Distance to the system. All values are based on \textit{Gaia} parallaxes. (3) Stellar mass (4) Stellar age (5) Planet orbital radius (6) Gas surface density at the planet's orbital radius, based on spatially resolved
C$^{18}$O emission from inside of the gap. The quoted range accounts for uncertainty in the CO:H$_2$ abundance. Ellipses mark systems without C$^{18}$O data.
(7) Midplane temperature of the circumstellar disc at the planet's orbital radius, estimated either by assuming the disc is passively heated by its host star or from fits to mm-wave observations. (8) Planet mass. In most cases the mass is estimated from the width of the gap (\protect \citealt{zhang_etal_2018}) and the quoted range corresponds to a Shakura-Sunyaev viscosity parameter of $\alpha = 10^{-5}$-$10^{-3}$. (9) Planet's time-averaged accretion rate $\dot{m}_{\rm p} = m_{\rm p}/t_{\rm age}$. \nickk{(10) Planet's radius $r_{\rm shock}$, calculated from the mass-radius relation in equation \ref{eqn:rshock} with $F_{\rm r} = 3$. Entries where the first number $>$ second number reflect how $r_{\rm shock}$ may decrease with $m_{\rm p}$.}
(11) Planet's accretion luminosity $L_{\rm acc} = Gm\dot{m}_{\rm p}/r_{\rm shock}$. 
(12) Temperature behind the accretion shock, equal to the value required for thermalised material to radiate $L_{\rm acc}$. (13) Wavelength at which the Planck spectrum $B_{\nu}(T_{\rm shock})$ (measured per unit frequency) peaks.
(14) Peak flux density of the planet, $F_{\nu} = B_{\nu}(T_{\rm shock}, \lambda_{\rm peak}) \pi r_{\rm shock}^2/d^2$, assuming no extinction. In most cases, the values for $t_{\rm age}$, $m_{\rm p}$, $\Sigma_{\rm csd}$, $T_{\rm csd}$, $\dot{m}_{\rm p}$, $r_{\rm shock}$, $\log_{10}L_{\rm acc}$, $T_{\rm shock}$, $\lambda_{\rm peak}$, and $F_{\nu}$ are rounded to one significant figure. Extra precision is given in cases where the lower and upper bounds are the same after rounding.}
\label{tab:compilation}
\end{table*}

\section{Summary and Discussion}
\label{sec:summary}

To aid ongoing efforts to directly image protoplanets, we have computed the broadband spectral energy distributions (SEDs) of planets embedded within and accreting from their parent circumstellar discs. The accreting protoplanet is modeled as a spherical blackbody powered by accretion. It gives rise to radiation 
a few microns in wavelength. The longer wavelength emission, from the mid-infrared to the radio, arises from circumplanetary dust whose distribution can be spherical or disc-like. The dust is heated externally by planetary radiation and internally by compression and/or viscous dissipation. 

A lesson learned from our modeling is that circumplanetary material is hard to see against the background circumstellar disc at wavelengths longer than $\sim$20 microns. The protoplanet has to contend with a typically flared circumstellar disc that intercepts an order-unity fraction of light from its host star. A patch of the circumstellar disc, located near the planet and having dimensions on the order of a scale height, can emit up to $\sim$0.1\% of the stellar luminosity. Most of that power emerges from the uppermost layer of the circumstellar disc which directly absorbs starlight at visible wavelengths, reprocessing it to the same long wavelengths characterizing circumplanetary material. Even the circumstellar gaps observed by the Atacama Large Millimeter Array (ALMA) may be optically thick to incident starlight, thereby emitting strongly from mid-infrared to radio wavelengths. Such contaminating light may frustrate searches for protoplanets using ALMA or the long-wavelength portion of the James Webb Space Telescope's Mid-Infrared Instrument (MIRI). 
\nick{Previous protoplanet SED models omitted consideration of background circumstellar disc light (e.g.~\citealt{zhu_2015}; \citealt{adams_batygin_2022}; \citealt{taylor_adams_2024}) or may not have resolved the circumstellar disc surface layers where starlight is absorbed (e.g. section 2.2 of \citealt{szulagyi_etal_sed2} and section 2.3 of \citealt{krieger_wolf_2022}).}



The good news is that protoplanets may outshine enshrouding circumstellar discs at shorter wavelengths because protoplanet temperatures of $\sim$300-1000 K (for objects in their final decade of radius contraction) are much greater than circumstellar disc temperatures of $\sim$30 K. Table \ref{tab:compilation} compiles protoplanets hypothesized to reside within gapped discs, and provides estimates for their accretion rates, luminosities, shock temperatures, and corresponding peak blackbody wavelengths and fluxes (without extinction). The brightest candidate sources all peak in the near-infrared, presenting potential targets for JWST’s Near-Infrared Camera (NIRCam) and the short-wavelength portion of MIRI. Searches using these instruments are in progress \citep[e.g.][]{cugno_etal_2023b}. We found that protoplanets more massive than Saturn could be detected using the NIRCam and MIRI coronographs. At these masses, near-infrared starlight scattered off the circumstellar disc is unlikely to introduce confusion (section \ref{subsec:csd}). Still, detection hinges on minimal circumplanetary dust extinction. It helps if circumplanetary dust has settled into a disc (Fig. \ref{fig:cpd_sed}). If circumplanetary material is more spherical (as might be expected for less massive protoplanets), dust must be depleted by about an order of magnitude compared to solar proportions to avoid obscuring the protoplanet (Fig. \ref{fig:cps_sed}). There is hope for subsolar dust-to-gas ratios inside of planet-carved gaps because of aerodynamic filtration \citep[e.g.][]{paardekooper_mellema_2004, dong_etal_2017}. 

\referee{Given the potential for confusion by thermal circumstellar disc emission at long wavelengths and scattered starlight at short wavelengths, and the threat of extinction from circumplanetary dust, it is perhaps no wonder that so few gap-opening protoplanets, many of which have suspected masses as low as $\sim$$0.03 m_{\rm J} \simeq 10 m_\oplus$, have not been detected with confidence. Exceptions include the confirmed protoplanets PDS 70b and c, which are super-Jupiters privileged to reside within an especially well-evacuated circumstellar disc cavity. To the ranks of PDS 70b and c we might hope to add the brightest sources in Table \ref{tab:compilation}: SR 4, Elias 24, MWC 758, HD 143006-G1, and AB Aur, whose masses are estimated to be comparable to if not greater than Jupiter's, and whose emission should be brighter than those of their local circumstellar discs at near-infrared (but not necessarily longer) wavelengths.}


\referee{In our modeling we have tried to be agnostic about circumplanetary geometry by considering both circumplanetary spheres and circumplanetary discs. Both geometries have been found in numerical simulations, depending on planet mass and how flow thermodynamics are treated. For low-mass subthermal planets (see equation \ref{eqn:bondi}), quasi-spherical envelopes have been observed in adiabatic (\citealt{fung_etal_2019}), isothermal (\citealt{choksi_etal_2023}), and radiation-hydrodynamic simulations \citep[][]{cimerman_etal_2017, moldenhauer_etal_2021}.
For more massive superthermal planets, both spheres and discs have been seen. Radiation-hydrodynamic simulations report nearly hydrostatic, spherical envelopes extending to a fraction of the Hill (not Bondi) sphere (\citealt{szulagyi_etal_2016,  lambrechts_etal_2019b, krapp_etal_2024}; but see section 4 of the latter work for a discussion of how this envelope structure might be an artefact of boundary conditions). 
Isothermal simulations in the superthermal regime produce more disc-like geometries within the Hill sphere \citep{choksi_etal_2023, li_etal_2023}. For the superthermal super-Jupiter protoplanets in PDS 70, disc-like geometries are supported by how well the near-infrared SED conforms to a relatively unobscured blackbody (Fig.~\ref{fig:pds70_sed}; see a similar no-extinction, disc scenario in section 3.3 of \citealt{choksi_chiang_2022}).}

A related theoretical question is how fast protoplanets accrete gas, and how this rate depends on planet and disc properties.
Recent work has established a hydrodynamical upper bound on accretion rates (\citealt{choksi_etal_2023}; \citealt{li_etal_2023}), but the extent to which actual accretion rates approach the theoretical maximum has yet to be calculated from first principles. At issue are mechanisms for angular momentum transport in the circumplanetary region; candidate mechanisms include a Lindblad torque from the host star (\citealt{zhu_etal_2016}; \citealt{xu_goodman_2018}), and self-gravity (\citealt{gammie_2001}). Although the case remains 
tenuous, PDS 70c may be accreting at nearly the maximum rate, predicted to be $1 \, m_{\rm J}/{\rm Myr}$ from equation (18) of \cite{choksi_etal_2023}. This rate is within a factor of two of the planet's time-averaged accretion rate $m_{\rm p}/t_{\rm age} \approx 0.6\,m_{\rm J}/\rm Myr$ (see also fig. 13 of \citealt{choksi_etal_2023}). It also yields an accretion luminosity of order $10^{-4} L_\odot$ that appears consistent with the observed near-infrared SED (\citealt{wang_etal_2020}) and the observed mm-wave flux \citep{benisty_etal_2021}, as we have shown with our SED model (Fig.~\ref{fig:pds70_sed}). Unexplained in this picture is why the mm excess detected around the sibling protoplanet PDS 70b, though similar in total flux to PDS 70c, is more spatially extended \citep{isella_etal_2019, benisty_etal_2021}.  Also puzzling is why an accretion-powered ultraviolet excess is observed for PDS 70b but not for PDS 70c  (\citealt{zhou_etal_2021}). Perhaps accretional excesses, like line emission for accreting protostars, are time-variable. 

\section*{Acknowledgements}
We thank Yuhiko Aoyama, Til Birnstiel, Brendan Bowler, Gabriele Cugno, Stefano Facchini, Gabriel-Dominique Marleau, Kees Dullemond, Gaspard Duchêne, Haochang Jiang, Jane Huang, Ya-Ping Li, Wenbin Lu, Ruth Murray-Clay, Eliot Quataert, Daniel Thorngren, Andrew Youdin, and Daniel Weisz for useful exchanges. \referee{The anonymous referee provided an insightful report that prompted us to consider compressional heating in our circumplanetary sphere models and led to other improvements.} Financial support was provided by NSF AST grant 2205500 and a Simons Investigator award, and an NSF Graduate Research Fellowship to NC.

\section*{Data availability}
Data and codes are available upon request of the authors.




\bibliographystyle{mnras}
\bibliography{planets_nick} 



\appendix

\section{CPS (Bondi) Density Profiles}
\label{appendix:bondi}
\referee{Our CPS density profiles are computed by numerically solving the continuity and momentum equations for steady, spherically symmetric accretion:
\begin{align}
\partial_r \rho_{\rm cps} &= \frac{2\rho_{\rm cps} }{ \gamma - 1 } \left( \frac{- 2/r - \partial_r \mathcal{M}/\mathcal{M} }{D} \right) \label{eqn:continuity} \\
\partial_r\mathcal{M} &= 
\frac{-\frac{Gm_{\rm p}}{r^2}\left(1 - \frac{2c_{\rm cps}^2r}{Gm_{\rm p}}\right) }{\mathcal{M}c_{\rm cps}^2 \left(1 - 1/D \right)\left(1 - \mathcal{M}^{-2}\right) } +  \frac{  2\mathcal{M} }{r \left(D - 1 \right) }  \label{eqn:mtm1}
\end{align}
for a gas that behaves adiabatically with index $\gamma$:
\begin{align}
c_{\rm cps} &= c_{\rm csd}\left(\frac{\rho_{\rm cps}}{\rho_{\rm csd} }\right)^{\frac{\gamma - 1}{2}} \,.\label{eqn:eos}
\end{align}
Here $\partial_r \equiv \partial / \partial r$,  $\mathcal{M}$ and $c_{\rm cps}$ are the local Mach number and sound speed, and $D \equiv 2/(\gamma-1) + 1$. Near the sonic point, equation (\ref{eqn:mtm1}) is difficult to integrate numerically because both the numerator and the denominator of the first term on the right-hand side vanish. We use L'Hôpital's rule to obtain an alternate expression valid in the limit $\mathcal{M} \rightarrow 1$:\footnote{ \referee{We take the solution with $\partial_r \mathcal{M} < 0$. The solution with $\partial_r \mathcal{M}  > 0$ describes a wind.  } }
\begin{align}
\partial_r \mathcal{M} &=  -\sqrt{-C/A} \label{eqn:mtm2} \\ 
A &= 2c_{\rm cps}^2 \mathcal{M} - 2c_{\rm cps}^2\mathcal{M}/D \nonumber \\ 
C &= -2Gm_{\rm p}\mathcal{M}^3/r^3 + 2c_{\rm cps}^2\mathcal{M}^3/r^2 + 8c_{\rm cps}^2\mathcal{M}^3/(r^2 D). \nonumber  
\end{align}
In our numerical integrations, we use (\ref{eqn:mtm2}) when $|\mathcal{M} - 1| < 10^{-5}$. We have checked that our numerical solution matches the analytic solution for $\gamma = 1$  \citep{cranmer_2004}.}

\section{Dust scattering in CPS models}
\label{appendix:scat}
\begin{figure*} 
\includegraphics[width=\columnwidth]{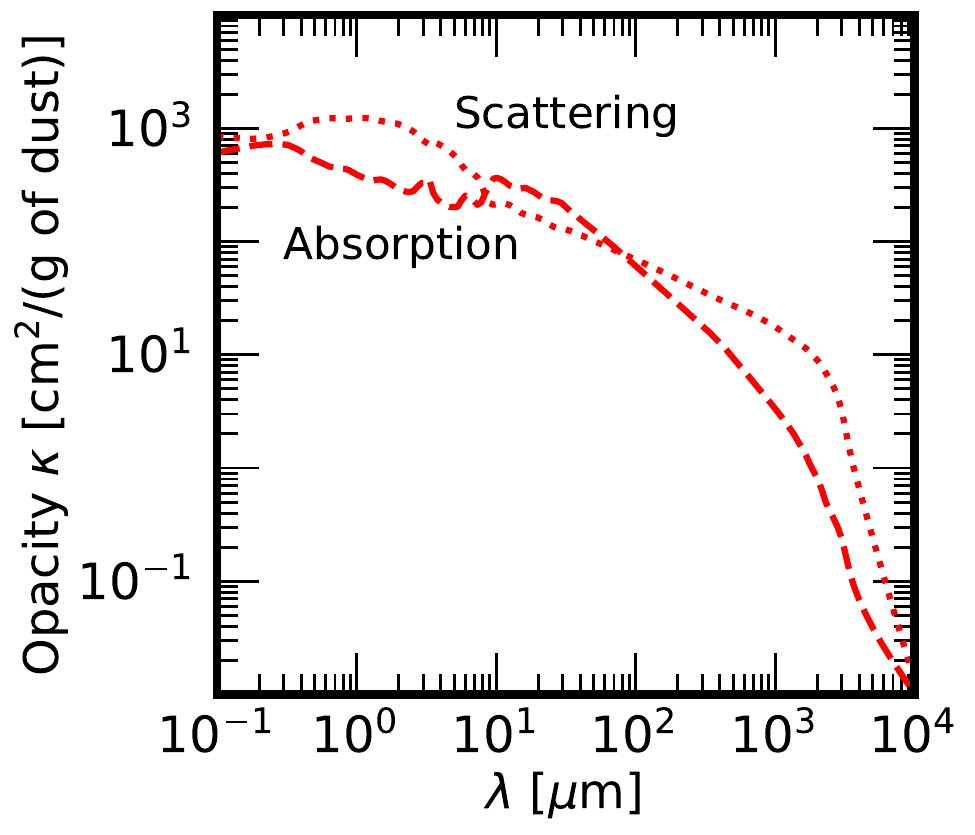}
\caption{ Dust scattering opacity (dotted curve) vs.~wavelength $\lambda$. These are computed using the \texttt{dsharp-opac} package for grains without water ice and have been averaged over the same grain size distribution used to calculate dust absorption opacities (dashed curve, same as in Fig. \ref{fig:opacity}) in the rest of this paper. }
\label{fig:opacity_scat}
\end{figure*}

\begin{figure*} 
\includegraphics[width=\columnwidth]{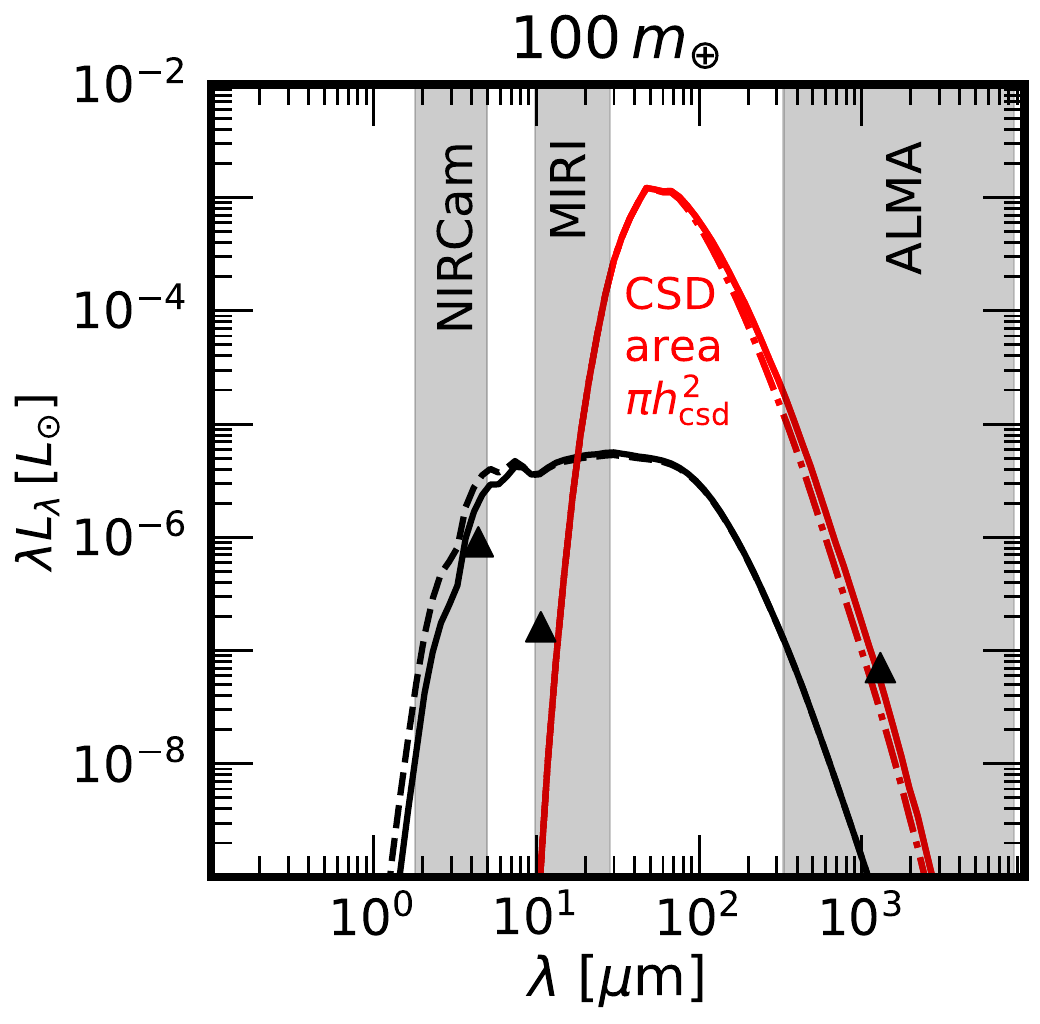}
\caption{ SEDs of CPS models with $\fdtg = 3 \times 10^{-3}$ around a 100$\Mearth$ planet. The solid black curve shows the emergent SED from the CPS when isotropic dust scattering is included, and the dashed curve shows the SED without scattering. 
}
\label{fig:cps_scat}
\end{figure*}

To assess the importance of light scattering by dust grains in our circumplanetary sphere (CPS) models, we run a $100\,m_\oplus$ CPS model with isotropic dust scattering included in $\radmc$. Figure \ref{fig:opacity_scat} 
plots the scattering opacity as a function of wavelength for the same grain size distribution we used to calculate absorption opacities. Figure \ref{fig:cps_scat} shows the emergent CPS SED with and without scattering. Including scattering reduces slightly the amount of planetary radiation that escapes the CPS. For a dust-to-gas ratio of $f_{\rm dust} = 3 \times 10^{-3}$, the reduction in brightness in the F444W and F1065C bandpasses is at most $\sim$10\%. 

\section{New Data in Table 2}
\label{appendix:data}
\nick{Table \ref{tab:compilation} includes entries for candidate protoplanets in MWC 758 and AB Aur which are not found in the corresponding table in \cite{choksi_etal_2023}. We describe here the data underlying these new entries. }

\subsection{MWC 758}

\cite{wagner_etal_2023} identified a near-infrared point source and potential protoplanet in the disc around the star MWC 758. The mass of the star is $M_{\star} = 1.9\,M_{\odot}$ and its age is $t_{\rm age} = 10\,\rm Myr$ (\citealt{garufi_etal_2018}; note that the latter value is larger than the pre-\textit{Gaia} value quoted in \citealt{wagner_etal_2023}). The point source has a projected separation from the star of $a = 100$ au \citep{wagner_etal_2023}. The mass range $m_{\rm p} = \left(3 - 100\right)\,m_{\rm J}$ quoted in Table \ref{tab:compilation} covers the range of masses for which \cite{wagner_etal_2023} obtained acceptable fits to the near-infrared photometry (their fig. 3). \cite{andrews_etal_2011} fit the mm-wave SED of the CSD and found $h_{\rm csd}/a = 0.18$ at the location of the planet. We estimate the local midplane temperature $T_{\rm csd}$ assuming the CSD is vertically isothermal and in hydrostatic equilibrium.

\subsection{AB Aur}
\nick{\cite{currie_etal_2022} identified a spatially resolved source at optical/near-infrared wavelengths in the AB Aur disc which they argue arises from an embedded protoplanet  (but see \citealt{zhou_etal_2022, zhou_etal_2023} for an alternate interpretation). We use the planet and system properties listed in their table 1. The mass of the star is $M_{\star} = 2.4\,M_{\odot}$ and its age is $t_{\rm age} = 3\,\rm Myr$. The source has a projected separation from the star of $a = 94$ au. The lower limit on the planet mass $m_{\rm p} = 9\,m_{\rm J}$ comes from modeling by \cite{currie_etal_2022} of the source's observed SED. The upper limit $m_{\rm p} = 130\,m_{\rm J}$ derives from \textit{Gaia} and \textit{Hipparcos} astrometry of the host star. We set $T_{\rm csd} = 30$ K, as determined by \cite{tannirkulam_etal_2008} from their fits to the infrared-to-mm CSD SED (their fig. 13).}

\bsp	
\label{lastpage}
\end{document}